\documentclass[twocolumn,amsmath,amssymb,floatfix,pre,floatfix]{revtex4}

\usepackage{graphicx,color}
\definecolor{brown}{rgb}{0.63,0.27,0.18}
\definecolor{orange}{rgb}{1.00,0.65,0.00}

\usepackage{afterpage} 

\usepackage{moresize}
\usepackage{dcolumn}
\usepackage{bm,multirow}

\makeatletter
\newcommand*{\balancecolsandclearpage}{%
  \close@column@grid
  \twocolumngrid
}
\makeatother

\newcommand{\be}{\begin{equation}}
\newcommand{\ee}{\end{equation}}
\usepackage{color}

\begin{document}

\newcommand {\rsq}[1]{\left< R^2 (#1)\right.}
\newcommand {\rsqL}{\left< R^2 (L) \right>}
\newcommand {\rsqbp}{\left< R^2 (N_{bp}) \right>}
\newcommand {\Nbp}{N_{bp}}
\newcommand {\etal}{{\em et al.}}
\newcommand{\Ham}{{\cal H}}
\newcommand{\AngeloComment}[1]{\textcolor{red}{(AR) #1}}
\newcommand{\AndreaComment}[1]{\textcolor{blue}{A: #1}}
\newcommand{\JanComment}[1]{\textcolor{green}{(JS) #1}}

\newcommand{\NewText}[1]{\textcolor{orange}{#1}}
\newcommand{\scs}{\ssmall}

\newcommand{\Tau}{\mathrm{T}}



\title{
Spatial organization of slit-confined melts of ring polymers with non-conserved topology: A lattice Monte Carlo study
}

\begin{abstract}
We present Monte Carlo computer simulations for melts of semiflexible randomly knotted and randomly concatenated ring polymers on the fcc lattice and in slit confinement.
Through systematic variation of the slit width at fixed melt density, we first explore the influence of confinement on single-chain conformations and inter-chain interactions.
We demonstrate that confinement makes chains globally larger and more elongated, while enhancing both contacts and knottedness propensities.
As for multi-chain properties, we show that ring-ring contacts decrease with the confinement, yet neighbouring rings are more overlapped as confinement grows.
These aspects are reflected on the decrease of the links formation between pairs of rings.
The results suggest that confinement can be used to fine-tune the mechanical properties of the polymer network.
In particular, confinement biases the synthesis of networks that are softer to mechanical stress.
Finally, in connection with a previous study of us and recent simulations on two-dimensional polymer melts, our findings suggest that entanglements in polymer melts arise from pairwise ring-ring links alone. 
\end{abstract}
\author{Mattia Alberto Ubertini}
\email{mubertin@sissa.it}
\affiliation{Scuola Internazionale Superiore di Studi Avanzati (SISSA), Via Bonomea 265, 34136 Trieste, Italy}
\author{Angelo Rosa}
\email{anrosa@sissa.it}
\affiliation{Scuola Internazionale Superiore di Studi Avanzati (SISSA), Via Bonomea 265, 34136 Trieste, Italy}
\date{\today}
\maketitle
\section{1. Introduction}\label{sec:Intro}
Recent years have witnessed a growing interest in the design of so called {\it smart} materials, such polycatenanes and polyrotaxanes~\cite{wu2017poly,hart2021material}, whose microscopic components are constituted by ring polymers interlocked to each other by topological links that can be artificially synthesised following precise chemical routes.
Interestingly similar devices can be also prepared by employing biological components, mainly DNA plasmid rings~\cite{krajina2018active} which interlock to each other through the action of the enzyme {\it topoisomerase-II} (TopoII) and form a molecular state termed {\it Olympic} hydrogel which has been first theorized by de Gennes in 1997~\cite{deGennes1997}.
Remarkably, similar molecules can be also found in Nature: a classical example is the kinetoplast DNA~\cite{chen1995topology} present in the mitochondria of certain {\it Trypanosoma} parasites.

Similarly to covalent bonds stabilizing the shape of a molecule, topological links remain stable at room temperature which guarantees the corresponding molecule to maintain a relatively well characterized spatial conformation.
On the other hand, since the single ring constituents are not rigid objects but they fluctuate~\cite{chiarantoni2022effect} as ordinary polymers typically do~\cite{DoiEdwardsBook,RubinsteinColbyBook}, these molecules display unusual mechanical properties under stress and tunable viscoelasticity that can be exploited in a wide number of practical applications (molecular machines and drug delivery~\cite{rauscher2020dynamics,rauscher2020thermodynamics}, to name a few), thus justifying the adjective ``smart'' employed for these materials.

The preparation of topological materials with well designed properties is a delicate balance between many parameters: indeed, several numerical studies~\cite{lang2012effect,lang2014swelling,ubertini2021computer,ubertini2023topological} have characterized the topological state of systems made up of randomly concatenated and knotted polymer rings, and have shown that the resulting networks can be controlled using experimentally tunable parameters such as the length of the polymer chain, the density of the polymer solution, and the bending stiffness of the polymer fiber.
So far though, {\it geometric confinement} as a way to drive the synthesis of concatenated ring networks has received considerable less attention.
Yet, recent experiments~\cite{soh2021equilibrium} performed on kinetoplast DNA~\cite{chen1995topology} at varying degree of {\it slit confinement} have foreseen the possibility of exploiting geometric constraints to bias the synthesis of a DNA-based network, similarly to the one discussed in Ref.~\cite{krajina2018active}. 

In this work, we explore how geometric constraints, under the form of slit confinement, can affect the structural properties of systems of strand-crossing rings.
To this purpose, we perform extensive dynamical simulations of highly entangled systems of randomly concatenated and knotted rings employing the kinetic Monte Carlo algorithm introduced by us~\cite{ubertini2021computer} for studying these systems at bulk conditions.
Varying the degree of confinement, we quantify its influence on the metric properties of the rings, which present interesting non-monotonous behaviour, as well as topological ones, in particular knotting probability is highly enhanced by reducing the height of the slit, while the linking between the rings is diminished.
These findings suggest that geometric confinement can be used as a powerful tool to control the topology of the resulting networks and their elastic properties. 

The paper is structured as the following.
In Section~1 
we present and discuss the Monte Carlo lattice polymer model, we introduce the notation and we explain how to detect and compute topological invariants for the characterization of knots and links in the system.
In Sec.~2 
we present the main results of our work, while in
Sec.~3 
we provide some discussion and conclusions regarding the role of slit confinement in shaping both single-chain and inter-chain properties of the resulting polymer networks.
Additional figures have been included in the Supporting Information (SI) file.

\section{2. Model and methods}\label{sec:ModelMethods}

\subsection{2.1. Polymer model}\label{sec:PolymerModel}
We consider polymer melts made of $M$ randomly concatenated and randomly knotted ring polymers of $N=320$ monomers each on the fcc lattice; the fcc unit step $a$ is taken as our unit length.
The simulations are based on the kinetic Monte Carlo (kMC) algorithm introduced by us in~\cite{ubertini2021computer}.
Since then, the algorithm has been variously applied to study melts of non-concatenated and unknotted rings~\cite{ubertini2022double} and the connection between entanglements and physical links in semiflexible chain melts~\cite{ubertini2023topological}.
In this article we limit ourselves to summarizing the essential details of the numerical protocol, while referring the reader to our past works for more details.

\begin{table}
\begin{tabular}{cccc}
\hline
\hline
\\
\, ${\rm H} / a$ \, & \, $\hat{{\rm H}}$ \, & \, $M$ \, & \, $\langle b \rangle / a$ \, \\
\hline
$2.12$ & $0.30$ & $420$ & $0.656$ \\
$3.53$ & $0.50$ & $422$ & $0.658$ \\
$4.95$ & $0.70$ & $420$ & $0.659$ \\
$6.36$ & $0.90$ & $427$ & $0.659$  \\
$7.78$ & $1.10$ & $420$ & $0.660$ \\
$10.61$ & $1.51$ & $420$ & $0.660$ \\
$13.43$ & $1.91$ & $422$ & $0.660$ \\
$17.68$ & $2.51$ & $430$ & $0.660$ \\
$20.51$ & $2.91$ & $433$ & $0.660$ \\
${\rm bulk}$ & -- & $420$ & $0.663$ \\
\hline
\hline
\end{tabular}
\caption{
Values of physical parameters for the ring polymer melts investigated in this paper.
$a$ is the unit distance of the fcc lattice and the monomer number per fcc lattice site is equal to $\frac54=1.25$, see text and Refs.~\cite{ubertini2021computer,ubertini2022double,ubertini2023topological} for details.
(i) ${\rm H}$, height of the slit.
(ii)  $\hat{\rm H} = \frac{{\rm H}}{\sqrt{\langle R^2_g \rangle_{\rm bulk}}}$, ratio between the height of the slit and the root-mean-square gyration radius of rings in bulk ({\it i.e.}, no confinement) conditions.
(iii) $M$, total number of simulated chains in the melt.
(iv) $\langle b \rangle$, mean bond length~\cite{MeanBondLengthNote}.
}
\label{tab:PolymerModel-LengthScales}
\end{table}

Essentially, the polymer model takes into account:
(i) chain connectivity,
(ii) bending stiffness,
(iii) excluded volume,
(iv) topological rearrangement of polymer chains.
Finally, and for the first time, in this work we consider (v) slit confinement in the model.
For the implementation of chain dynamics, the following combination of MC moves -- that automatically take into account excluded volume interactions -- are used:
\begin{itemize}
\item[(a)]
Topology-{\it preserving} moves (termed {\it Rouse-like} and {\it reptation-like}, see~\cite{ubertini2021computer}) that automatically enforce excluded volume interactions.
By construction, these moves enable two (and no more than two) consecutive bonded monomers along each single chain to occupy the same lattice site: by allowing to store contour length along the polymer filament, this numerical ``trick'' makes the chains locally elastic and facilitates global chain equilibration.
Because of that, the bond length is a fluctuating quantity with mean value $= \langle b \rangle$: in particular, the latter is insensitive to confinement (the measured values for $\langle b\rangle$ are reported in Table~\ref{tab:PolymerModel-LengthScales}).
In this way, the mean polymer contour length is $L = N \langle b\rangle$ and, similarly, the mean contour length of a sub-chain of $n$ monomers is $\ell =n\langle b\rangle$.
\item[(b)]
Topology-{\it changing} moves~\cite{ubertini2021computer} that induce random strand crossings between nearby polymer filaments at a tunable rate: we set this rate to $10^4$ kMC elementary steps, consistent with our previous works~\cite{ubertini2021computer,ubertini2022double,ubertini2023topological}.
Strand-crossings between filaments of the same ring can result in the creation or destruction of knots, while inter-ring crossings may cause either catenation or decatenation.
The model has been shown to exhibit dynamical behavior consistent with the experiments~\cite{krajina2018active}, specifically dynamic ``fluidization'' of the rings due to topological violations through strand-crossings.
Thus, by performing simulations of strand-crossing rings, we sample the ensemble of the network structures formed by randomly concatenated and knotted rings at the given density and in slit confinement (see below for details).
\end{itemize}
Then, bending stiffness is modelled in terms of the Hamiltonian (in Boltzmann units, $\kappa_BT$):
$$
\frac{\mathcal H}{\kappa_B T} = -\kappa_{\rm bend} \! \sum_{i=1}^{N \! \langle b\rangle /a} \! \cos\theta_i \, ,
$$
where $\kappa_{\rm bend} = 2$ is the bending stiffness and $\theta_i$ is the angle between consecutive bonds along the chain, with periodic conditions -- due to ring geometry -- assumed  for all the chains.
By fixing the monomer number per fcc lattice site equal to $\frac54=1.25$~\cite{ubertini2021computer,ubertini2022double,ubertini2023topological}, the chosen bending stiffness corresponds to the chain Kuhn segment $\ell_K / a = 3.4$~\cite{ubertini2022double} which is high enough to guarantee that distinct polymers are in an effective highly entangled state.

Finally, ring polymers are subject to slit confinement.
This particular form of constraint is imposed by forcing the chains to move on the fcc lattice, with periodic boundary conditions on the $xy$-plane and hard boundaries in the $z$-direction placed in $z = 0$ and $z = {\rm H}$.
We vary the height of the box ${\rm H}$ to study different confinement regimes, while adjusting the lateral box sides $L_x = L_y$ to keep density constant.
The degree of confinement is quantified by the ratio $\hat{\rm H} = {\rm H} / \sqrt{\langle R^2_g\rangle_{\rm bulk}}$, expressing the ratio between the height, or width, of the slit $\rm H$ and the root-mean-square gyration radius (see definition~\eqref{eq:GyrationRadius}), $\sqrt{\langle R^2_g\rangle_{\rm bulk}} / a = \sqrt{49.66}$~\cite{ubertini2022double}, of rings in bulk conditions.
We investigate system's behavior from highly confined ($\hat{\rm H} \simeq 0.30$) to mildly confined ($\hat{\rm H} \simeq 2.91$) regimes and systematically compare the results with the corresponding values in bulk.
Wherever appropriate, we have also compared the systems here with melts of unknotted and non-concatenated rings in bulk~\cite{ubertini2022double}.
We simulate $M \simeq 420$ chains, comprising a total of $N\times M \simeq 134400$ monomers, with $M$ slightly adjusted to maintain a constant density (see Table~\ref{tab:PolymerModel-LengthScales} for specific numbers).

To assess meaningful chain statistics and as in our other works~\cite{ubertini2021computer,ubertini2023topological} on similar polymer systems, we run simulations long enough in order to get properly equilibrated melts.
This is visualized in Figure~S1 
in SI, that shows plots of the monomer time mean-square displacement in the frame of the centre of mass of the corresponding chain (the so called $g_2$~\cite{KremerGrest1992}) as a function of the MC simulation time $\tau_{\rm MC}$.
As known, provided long-enough simulations are available, $g_2$ displays a plateau that is indicative of the equilibration of the system.
All our systems display corresponding plateaus, that demonstrates that equilibration has been reached for all the cases considered.
Accordingly, the time scale to reach the corresponding plateau corresponds to the portion of the trajectory that has been discarded from the computation of the relative observables.

\subsection{2.2. Detection of knots and links}\label{sec:KnotsLinksDetection}
In order to characterize the topological states of the rings in the melt, we follow closely the pipeline recently developed by us~\cite{ubertini2023topological}.
Specifically, we employ a numerical algorithm which ``shrinks'' or simplifies each ring to its ``primitive'' shape, {\it i.e.} without violating topological constraints: in this way we detect knots and links at any order, {\it i.e.} pairwise links as well as three-chain links like the {\it Borromean} ring configuration $6^3_2$ (see Sec.~2.3). 
The algorithm is able to return the irreducible knotted or linked structure which we further characterize by computing their topological invariants.
For knots, in particular, we compute the corresponding Jones polynomial~\cite{Jones1985} using the Python package {\it Topoly}~\cite{dabrowski2021topoly}. 
Instead, for two-body links we compute the Gauss linking number (GLN): 
\begin{equation}\label{eq:DefineGLN}
{\rm GLN} \equiv \frac1{4\pi} \oint_{{\mathcal C}_1} \oint_{{\mathcal C}_2} \frac{(\vec r_2 - \vec r_1) \cdot (d{\vec r}_2 \wedge d{\vec r}_1)}{|\vec r_2 - \vec r_1|^3} \, ,
\end{equation}
which gives the number of times two closed loops $\mathcal C_1$ and $\mathcal C_2$, parametrized respectively by coordinates $\vec r_1$ and $\vec r_2$, wind around each other. 
While unconcatenated rings have $\rm{GLN}=0$, it is known that there exists concatenated pairs with $\rm{GLN}=0$ (for instance, the so called {\it Whitehead} link configuration $5^2_1$).
In these ``pathological'' cases, the ones detected via our shrinking algorithm have been successively identified by computing the Jones polynomial using {\it Topoly} again.
We compute the Jones polynomials also for three-body irreducible links (for instance, Borromean rings) where a pairwise topological invariant such as the ${\rm GLN}$ fails (see Sec.~3.2.2). 

\subsection{2.3. Notation}\label{sec:Notation}
As for rings' metric properties, for some observables $\mathcal{O}$ which can be expressed as a function of monomers' coordinates we study separately the contributions $\mathcal{O}_{\perp}$ and $\mathcal{O}_{\parallel}$, respectively perpendicular (or, transverse) and parallel to the plane of the slit (which, by construction (see Sec.~2.1), 
coincides with the $xy$-plane).

As for rings' topological properties, in referring to a given knot or link we employ the conventional notation illustrated in the book by Rolfsen~\cite{Rolfsen2003KnotsAL}.
Namely, a knot or a link is defined by the symbol $K_i^p$ where:
$K$ represents the number of irreducible crossings of the knot (or the link),
$p$ is the number of rings which takes part in the topological structure ({\it e.g.}, $p=2$ for two-chain links) and $i$ is an enumerative index assigned to distinguish topologically {\it non-equivalent} structures having the same $K$ and $p$.

\section{3. Results}\label{sec:Results}

\subsection{3.1. Single-chain properties}\label{sec:SingleChainProperties}

\subsubsection{3.1.1. Rings' size and shape}\label{sec:RingsSizeShape}
First, we characterize the impact of slit confinement on the size and shape of the rings.
To this purpose, for each ring of the system we compute the $3 \times 3$ symmetric gyration tensor $Q_{\alpha\beta} = Q_{\beta\alpha}$ ($\alpha, \beta = x, y, z$) defined as:
\begin{equation}\label{eq:Gyration_tensor}
Q_{\alpha\beta} = \frac1N \sum_{m=1}^{N} \left( r_{m,\alpha} - r_{{\rm CM},\alpha} \right) \left( r_{m,\beta} - r_{{\rm CM},\beta} \right) \, ,
\end{equation}
where $r_{m, \alpha}$ is the $\alpha$-th Cartesian component of the spatial position $\vec r_m$ of monomer $m$ and $\vec r_{\rm CM} \equiv \frac1N \sum_{m=1}^N \vec r_m$ is the center of mass of the chain.
The mean values of the eigenvectors of $Q$ ordered in descending order, $\langle \lambda^2_1 \rangle \geq \langle \lambda^2_2 \rangle \geq \langle \lambda^2_3 \rangle$, quantify the mean spatial elongations of the polymers on the corresponding principal axes, while the mean value of the trace of $Q$, $\langle {\rm tr}Q \rangle = \sum_{\alpha=1}^3 \langle \lambda_{\alpha}^2 \rangle$, is equal to the mean-square gyration radius or size,
\begin{equation}\label{eq:GyrationRadius}
\langle R_g^2 \rangle \equiv \frac1N \sum_{m=1}^N \langle \left({\vec r} _m - {\vec r}_{\rm CM} \right)^2\rangle = \langle {\rm tr}Q \rangle = \sum_{\alpha=1}^3 \langle \lambda_{\alpha}^2 \rangle \, ,
\end{equation}
of the chain.

\begin{figure*}
\includegraphics[width=1.00\textwidth]{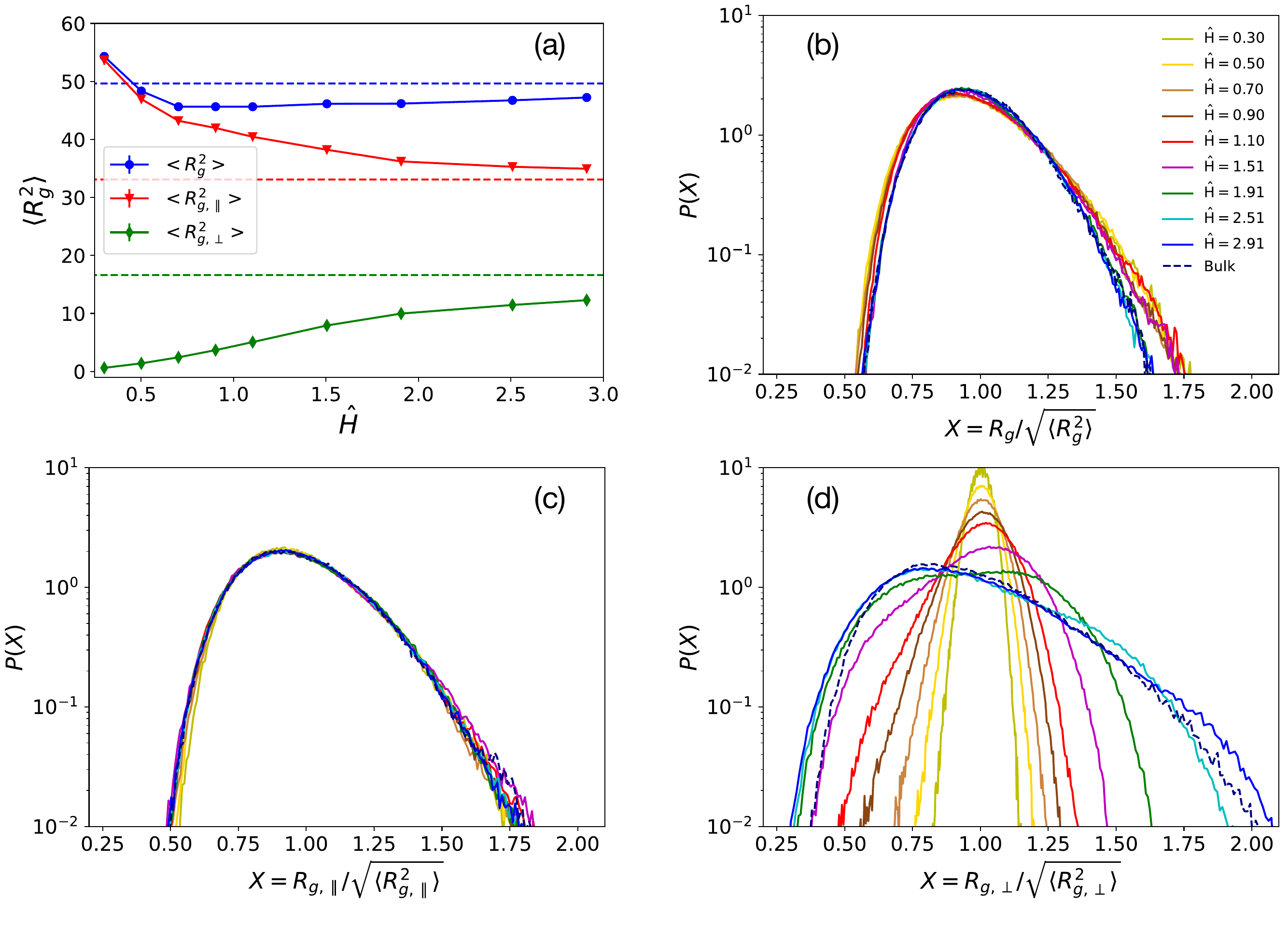}
\caption{
(a)
Ring mean-square gyration radius ($\langle R_g^2 \rangle$) with its parallel ($\langle R_{g, \parallel}^2 \rangle$) and transverse ($R_{g,\perp}^2$) components as a function of the degree of confinement $\hat{\rm H}$ (see Sec.~2.1 
for definition).
The dashed lines are for the values of the bulk system ({\it i.e.}, no confinement). Error bars are smaller than the symbols size.
(b, c, d)
Scaling plots for, respectively, distribution functions of the ring gyration radius ($P ( R_g / \sqrt{\langle R_g^2 \rangle} )$) and of its parallel ($P ( R_{g, \parallel} / \sqrt{\langle R_{g, \parallel}^2 \rangle} )$) and transverse ($P ( R_{g, \perp} / \sqrt{\langle R_{g, \perp}^2 \rangle} )$) components, at different degrees of confinement $\hat{\rm H}$ (see legend in panel (b)).
The dashed line in each panel corresponds to the reference distributions in bulk conditions.
}
\label{fig:Rg}
\end{figure*}

The results for $\langle R_g^2 \rangle$ (Eq.~\eqref{eq:GyrationRadius}) and the perpendicular and parallel components, $\langle R_{g, \perp}^2\rangle$ and $\langle R_{g, \parallel}^2\rangle$, are reported in Fig.~\ref{fig:Rg}.
As ${\rm H}$ decreases, the transverse component $\langle R_{g, \perp}^2 \rangle$ decreases (green curve in Fig.~\ref{fig:Rg}(a)) as expected.
Conversely, the parallel component $\langle R_{g, \parallel}^2 \rangle$ grows with confinement (red curve in Fig.~\ref{fig:Rg}(a)) because the ring is forced to spread along the plane of the slit.
Together, these two effects produce a characteristic non-monotonic behavior in the overall $\langle R_g^2(\hat{\rm H}) \rangle$ (blue curve in Fig.~\ref{fig:Rg}(a)) with the minimum attained around $\hat {\rm H} \simeq 0.7$, {\it i.e.} where confinement effects are expected to become more pronounced.
Interestingly, for high confinement ($\hat{\rm H} \lesssim 0.3$), the rings are markedly larger than the bulk reference (blue dotted curve in Fig.~\ref{fig:Rg}(a)).
In a previous study~\cite{d2017linking} of randomly concatenated rings under slit confinement, the non-monotonic behavior was also observed but the swelling compared to the bulk state was not seen.
We attribute this discrepancy to the fact that, in the previous work, rings without excluded volume were considered which could have favored more compact conformations.

Beyond average values, we have also computed the corresponding probability distributions, $P(R_g)$, $P(R_{g,\perp})$ and $P(R_{g,\parallel})$, and represented each of them (see Fig.~\ref{fig:Rg}, panels (b) to (d)) in the corresponding scaled variable to ease comparison.
While the distributions of the parallel component of the gyration radius are fundamentally unaffected by confinement (Fig.~\ref{fig:Rg}(c)), the ones of the normal components (see Fig.~\ref{fig:Rg}(d)) undergo a significant change in shape as the confinement becomes stronger, in particular becoming more peaked.
Together these changes produce an interesting effect on the distributions of the full gyration radius (Fig.~\ref{fig:Rg}(b)), which are characterized by higher tails for the systems under confinement.
This suggests that, under confinement, rings assume more heterogeneous sizes.

\begin{figure}
\includegraphics[width=0.50\textwidth]{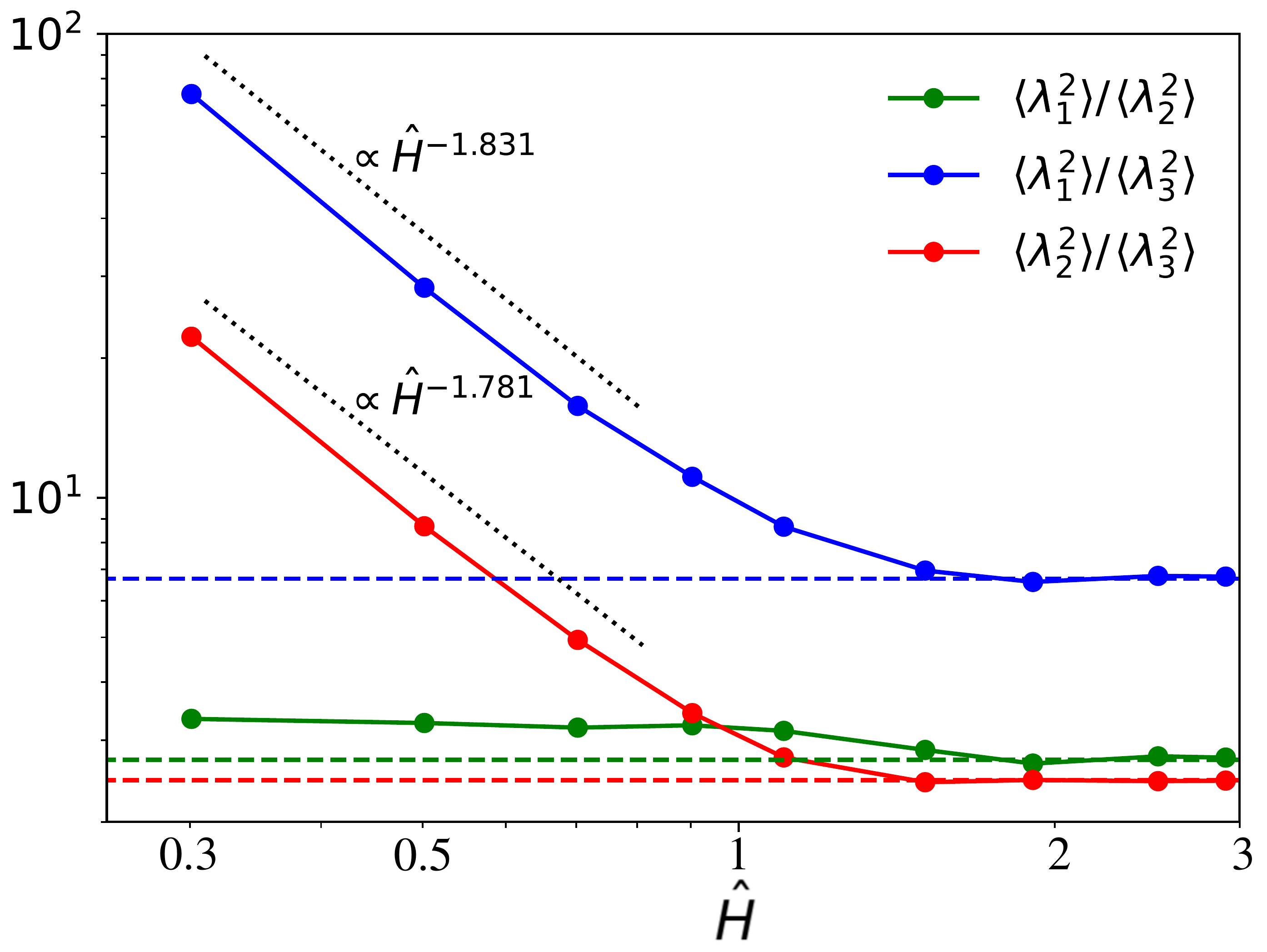}
\caption{
$\langle \Lambda_1^2 \rangle / \langle \Lambda_1^3 \rangle$ and $\langle \Lambda_2^2 \rangle / \langle \Lambda_1^3 \rangle$, average ring shapes expressed as the ratios between the two largest eigenvalues of the ring mean gyration tensor $Q$ (Eq.~\eqref{eq:Gyration_tensor}) and the smallest one, as a function of the degree of confinement $\hat{\rm H}$ (see Sec.~2.1 
for definition).
Dotted lines correspond to the power-law best fits obtained on the first three points of each curve.
Dashed horizontal lines correspond to the bulk reference values of the two ratios.
}
\label{fig:Shapes}
\end{figure}

We study then rings' shapes and anisotropies by looking at the ratios:
(i) $\langle \lambda^2_1 \rangle / \langle \lambda^2_2 \rangle$,
(ii) $\langle \lambda^2_1 \rangle/\langle\lambda^2_3 \rangle$
and
(iii) $\langle \lambda^2_2 \rangle/\langle \lambda^2_3 \rangle$.
The first ratio indicates the elongation or ``asphericity'' of the ring mean shape, while the other two measure the extent to which rings become effectively flat due to slit confinement.
Results are shown in Fig.~\ref{fig:Shapes}, where it is clear that at mild confinement $\hat{\rm H} \gtrsim 1.5$ rings attain the shame shape of the bulk ones.
At higher confinement ($\hat{\rm H} \lesssim 0.7$), the ratios to the smallest eigenvalues (blue and red curves in Fig.~\ref{fig:Shapes}) are described by characteristic power-law behaviors $\sim\!\hat{\rm H}^{-\alpha}$ with similar $\alpha$'s:
more precisely, the exponent for $\langle \lambda^2_1 \rangle/\langle\lambda^2_{3}\rangle$, $\alpha = 1.831\pm 0.001$, is only slightly larger than the exponent for $\langle \lambda^2_2 \rangle / \langle\lambda^2_3\rangle$, $\alpha \simeq 1.781 \pm 0.002$.
This difference is also evident in the behavior of $\langle \lambda^2_1 \rangle/\langle\lambda^2_2\rangle$ (green curve in Fig.~\ref{fig:Shapes}), which increases slightly with confinement.
In summary, our analysis shows that rings' flattening due to confinement biases the chains towards more elongated shape.

\subsubsection{3.1.2. Bond-vector correlation function}

%
\begin{figure*}
\includegraphics[width=1.0\textwidth]{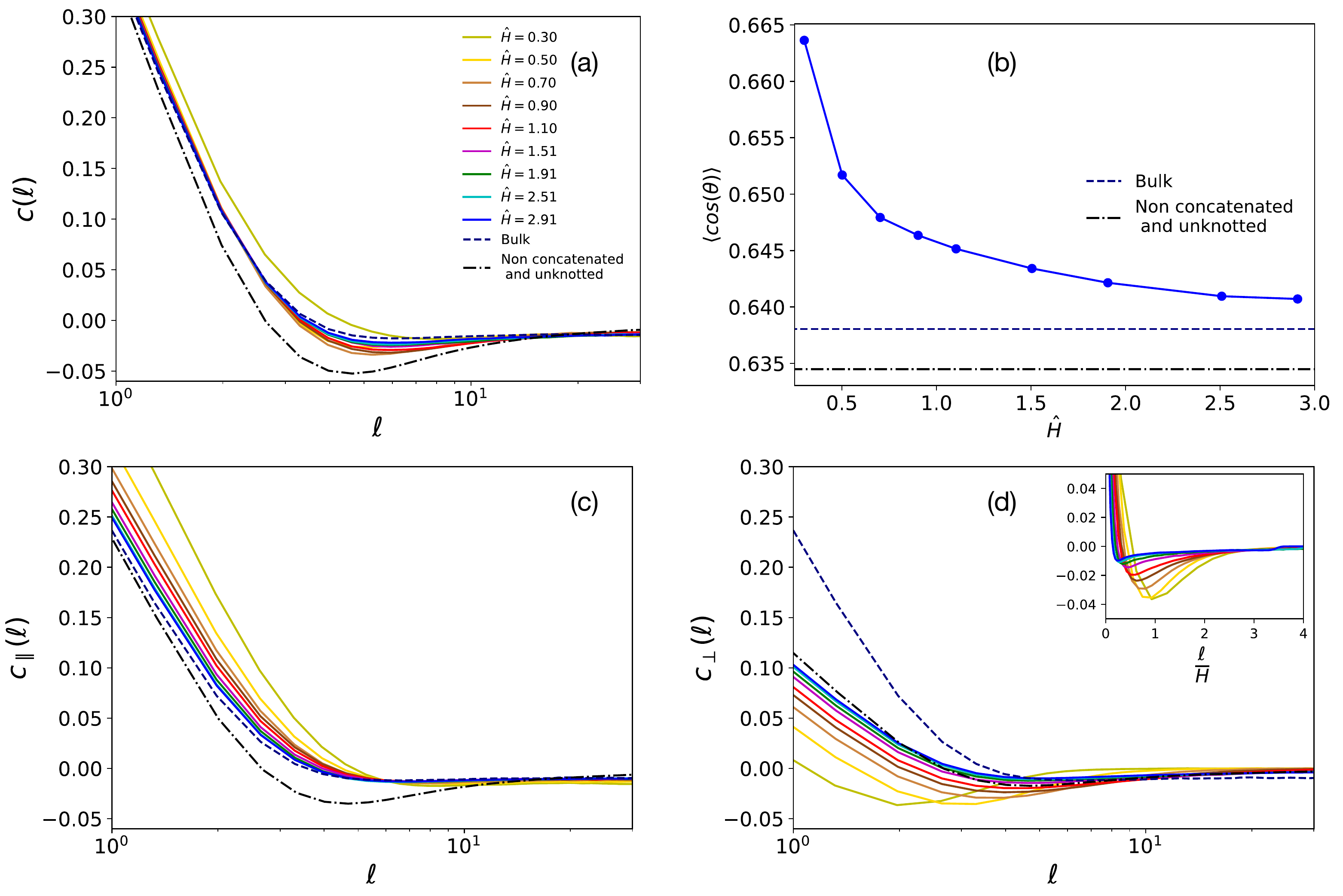}
\caption{
(a)
$c(\ell)$, bond-vector correlation function as a function of the contour length distance $\ell$.
Colors are for different confinements, dashed and dot-dashed lines are for bulk melts and melts of non-concatenated and unknotted rings (see legend).
(b)
$\langle \cos(\theta) \rangle$, mean cosine value between two consecutive bonds along the chain as a function of the degree of confinement $\hat{\rm H}$.
(c)
$c_{\parallel}(\ell)$, contribution to the bond-vector correlation function in the $xy$-plane parallel to the slit.
(d)
$c_{\perp}(\ell)$, contribution to the bond-vector correlation function orthogonal to the plane of the slit; in the inset the same quantity is represented as a function of the ring contour length normalized by the slit thickness, $\ell / H$.
Colors and symbols in panels (c) and (d) are as in panel (a).
}
\label{fig:Tangent}
\end{figure*}

We investigate now in more detail how the folding of polymer chains is affected by confinement by looking at the bond-vector correlation function,
\begin{equation}
c(\ell) \equiv \frac{\langle \vec{t}(\ell') \cdot \vec{t}(\ell'+\ell)\rangle}{\langle t(\ell')^2\rangle} \, ,
\end{equation}
as a function of the polymer contour length $\ell$.
This quantity gives useful insight when applied to bulk $3d$ melts of unknotted and non-concatenated rings, in particular its distinct~\cite{ubertini2022double} anti-correlation is a symptom of the double folding of the polymer chains at the entanglement scale (dot-dashed line in Fig.~\ref{fig:Tangent}(a)).
In contrast (dashed line in Fig.~\ref{fig:Tangent}(a)), bulk $3d$ melts of randomly knotted and concatenated rings exhibit normal exponential decay behavior~\cite{ubertini2023topological} and are not characterized by double folding, hence the anti-correlation is absent.

To investigate the impact of confinement on chain folding, we have computed $c(\ell)$ for the confined rings.
Results (Fig.~\ref{fig:Tangent}) exhibit several noteworthy effects.
Firstly (Fig.~\ref{fig:Tangent}(a)), for confined rings at small $\ell$ $c(\ell)$ decays more slowly than the bulk counterpart.
This is the consequence (Fig.~\ref{fig:Tangent}(b)) of the increase of the mean-cosine of the angle between consecutive bond vectors, $\langle \cos(\theta) \rangle$, as confinements increases: in other words, confined rings are slightly stiffer than the bulk reference and this confinement-enhanced stiffness grows with the confinement.
At the same time, $c(\ell)$ develops a characteristic anti-correlation that exhibits non-monotonic dependence on $\hat{\rm H}$: in particular the deepest minimum occurs at $\hat{\rm H} \simeq 0.7$, {\it i.e.} the same value at which the gyration radius (Fig.~\ref{fig:Rg}(a)) attains its minimum value.
Moreover, the minimum itself disappears at the highest level of confinement.
This peculiar behavior can be explained by considering the individual contributions of the parallel and transverse components of $c(\ell)$.
$c_{\parallel}(\ell)$ does not exhibit any minima (Fig.~\ref{fig:Tangent}(c)), while $c_{\perp}(\ell)$ displays a minimum for all values of $\hat{\rm H}$ (Fig.~\ref{fig:Tangent}(d)).
The mismatch in the values of $\ell$, at which $c_{\perp}(\ell)$ is minimum while $c_{\parallel}(\ell)\simeq 0$, causes the non-monotonicity of the full $c(\ell)$.
The latter goes to zero for similar values of $\ell$ for all $\hat{\rm H}$, demonstrating that correlations grow mildly with the confinement.
In contrast, $c_{\perp}(\ell)$ shows a minimum for $\ell$ close to the thickness of the slit ${\rm H}$ (Fig.~\ref{fig:Tangent}(d), inset).
This is due to the back-folding of the polymer filaments induced by the hitting with the impenetrable walls of the slit: of course this effect is more pronounced under strong confinement conditions, {\it i.e.} for ${\rm H} / \ell_K \leq 1$. 
Thus the minima in $c(\ell)$ appear when ${\rm H}$ has similar value to the correlation length of $c_{\perp}(\ell)$, indicating the competition between these two length scales.

\subsubsection{3.1.3. Contact probability}\label{sec:ContactProb}
As just shown, confinement alters the metric properties of the polymers.
Then, it is natural to expect that the consequent reorganization of the chains modifies the intra-chain polymer interactions.
To test this hypothesis, we compute the mean contact probability between two monomers at contour length separation $\ell = n\langle b\rangle$,
\begin{equation}\label{eq:ContactProb}
\langle p_c(\ell) \rangle = \left\langle \frac1N \sum_{i=1}^N \Theta( r_c - | {\vec r}_i - {\vec r}_{i+n} | ) \right\rangle \, ,
\end{equation}
where $\Theta(x)$ is the Heaviside step function and the ``contact distance'' $r_c$ is set to the unit lattice size $a$ (notice also that periodic conditions due to the ring geometry are tacitly assumed in Eq.~\eqref{eq:ContactProb}).

\begin{figure}
\includegraphics[width=0.50\textwidth]{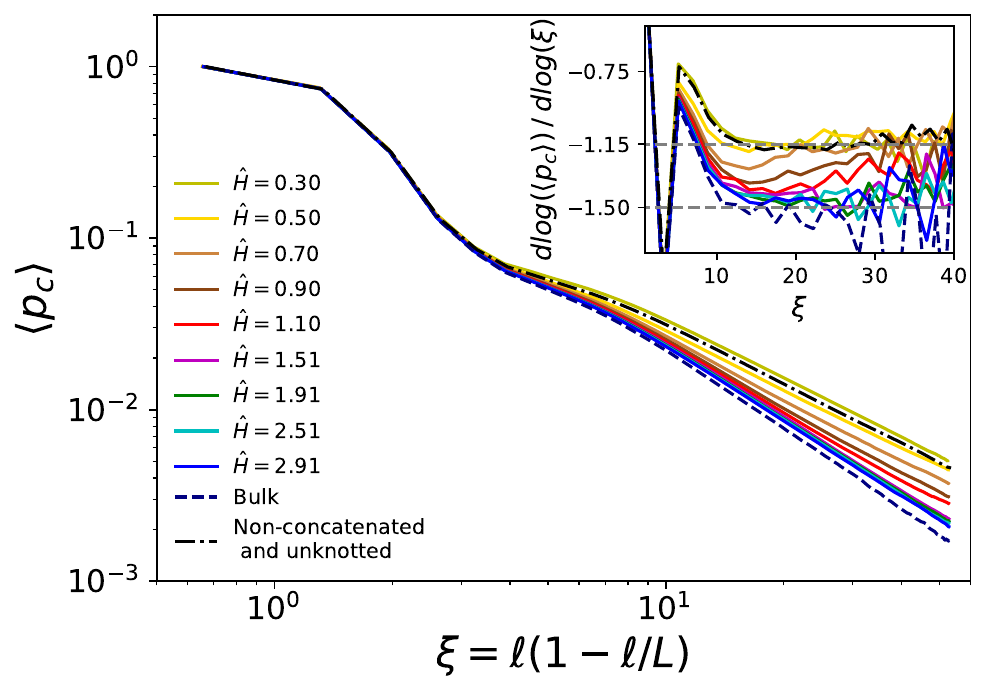}
\caption{
Mean contact probabilities, $\langle p_c \rangle$ (Eq.~\eqref{eq:ContactProb}), as a function of $\xi = \ell \left(1-\ell/L \right)$ where $\ell$ is the contour length separation between monomers and $L$ is ring total contour length.
Colors are for different confinements, dashed and dot-dashed lines are for bulk melts and melts of non-concatenated and unknotted rings (see legend).
Inset: local differential exponent $\equiv \frac{d\log\langle p_c \rangle}{d\log\xi}$.
}
\label{fig:Pc}
\end{figure}

Results are shown in Fig.~\ref{fig:Pc}, where $\langle p_c \rangle$ is plotted against the ``effective'' variable $\xi = \ell (1-\ell/L)$ in order to reduce~\cite{rosa2019conformational} finite-size effect due to the ring geometry.
First, one can notice that in bulk systems, as we let rings perform strand crossings, long-distance contacts decrease (dashed line) with respect to melts of non-concatenated and unknotted rings (dot-dashed line).
In contrast, confinement leads to an increase in the tail of the mean contact probability compared to the bulk reference.
Notably, at $\hat {\rm H} = 0.30$, the tail's slope is slightly less steep than in the non-concatenated state. 

To get more insight, it is interesting to look at the exponent controlling the asymptotic power-law decay, $\langle p_c \rangle \simeq \xi^{-\gamma}$ (Fig.~\ref{fig:Pc}, inset).
In bulk, strand-crossing rings attain ideal statistics characterized by $\gamma \simeq 1.5$, as confirmed by our previous findings~\cite{ubertini2021computer}.
In contrast, confinement leads to a decrease in $\gamma$ which becomes close to the same asymptotic value as the non-concatenated state, $\gamma \simeq 1.15$. 
Based on mean-field arguments~\cite{halverson2014melt}, $\gamma = d \nu$ where $d$ is space dimension and $\nu$ is the metric exponent of the chain relating~\cite{DoiEdwardsBook,RubinsteinColbyBook} the chain mean linear size to the number of monomers ({\it i.e.}, $\langle R^2_g \rangle \sim N^{2\nu}$).
Strand-crossing rings in bulk exhibit ideal statistics with $\nu = 1/2$~\cite{ubertini2021computer} and they are characterized by $\gamma = \frac32$ in three dimensions.
In confined systems, however, the rings cannot fold freely in three dimensions, effectively reducing the dimensionality of the system and resulting in a decrease in $\gamma$.

\subsubsection{3.1.4. Knots statistics}\label{sec:KnotsStatistics}
In our kMC algorithm two filaments from the same chain can cross and this event may induce the formation of a knot along the chain.
Characterization of knots spectra in confined systems have been addressed so far mostly for isolated chains~\cite{tesi1994knot,micheletti2011polymers,micheletti2012numerical}, while less results are available for confined systems at melt conditions.

\begin{figure}
\includegraphics[width=0.50\textwidth]{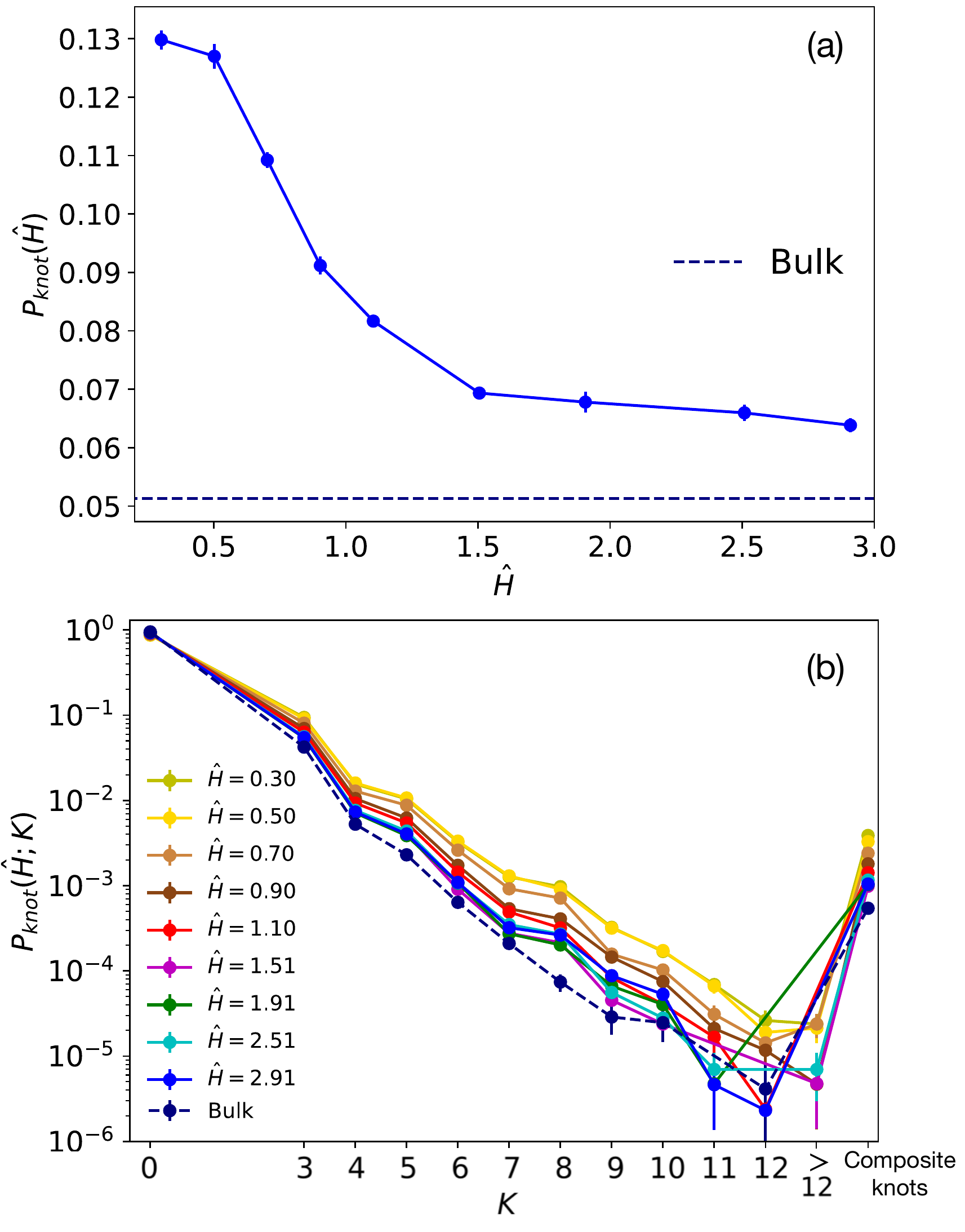}
\caption{
(a)
$P_{\rm knot}(\hat{\rm H})$, ring knotting probability (Eq.~\eqref{eq:KnottingProb}) for knots with $K$ crossings and as a function of the degree of confinement $\hat{\rm H}$.
The dashed line correspond to the value for the bulk melt.
(b)
$P_{\rm knot}(\hat{\rm H}; K)$, probability of finding a knot with crossing number $K$.
Colors are for different confinements, the dashed is for bulk melts (see legend).
$K=0$ correspond to the unknot and $P_{\rm knot}(\hat{\rm H}; K=0) = 1-P_{\rm knot}(\hat{\rm H})$ is its corresponding probability.
Knots with $>12$ crossings cannot be distinguished by {\it Topoly}~\cite{dabrowski2021topoly}.
Composite knots are knots made up by $2$ or more irreducible knots.  
}
\label{fig:Knots}
\end{figure}

To fill this gap, we have investigated the occurrence of knots by computing the Jones polynomial of each ring of our systems and, for simplicity, we present our results based on the number of irreducible crossings (denoted by $K$, see Sec.~2.3). 
Specifically, we have computed the probability, $P_{\rm knot}(\hat{\rm H}; K)$, of finding a knot with $K$ irreducible crossings at given confinement degree $\hat{\rm H}$ and the {\it cumulative} knotting probability:
\begin{equation}\label{eq:KnottingProb}
P_{\rm knot}(\hat{\rm H}) = \sum_{K=3}^{\infty} P_{\rm knot}(\hat{\rm H}; K) \, ,
\end{equation}
which gives the probability that a ring in the melt contains a knot (of any type).
As shown in Fig.~\ref{fig:Knots}(a), $P_{\rm knot}(\hat{\rm H})$ grows with the confinement and reaches the maximum value of $\simeq 0.13$ for the smallest $\hat{\rm H}$, resulting in an increase of $\simeq 130\%$ compared to bulk reference (dashed line).
Both in bulk and in confinement, the most common knot type is the simplest one, namely the {\it trefoil} knot $3_1$.
Overall (Fig.~\ref{fig:Knots}(b)), more complex knots are much less probable for all $\hat{\rm H}$ values, yet their abundance increases with confinement, see Fig.~\ref{fig:Knots}(b) for $P_{\rm knot}(\hat{\rm H}; K)$ and Fig.~S2 
in SI for the relative population of knot types with $K$ crossings. 
In conclusion, our analysis points out that confinement enhances the probability of knot formation, yet the overall occurrence of knots ({\it i.e.}, $P_{\rm knot}$) remains relatively low ($\lesssim 0.13$). 

\subsection{3.2. Chain-chain correlations}\label{sec:InterChainStatistics}

\subsubsection{3.2.1. Chain neighbours}\label{sec:Neighbors}
The increase of the long-range intra-chain contacts seen in Fig.~\ref{fig:Pc} may be indicative of the fact that confinement reduces the overlap between distinct chains or, in other words, ring-ring contacts should decrease. 
To test this hypothesis, we introduce the variable for the number of neighbors of ring $i$ $(i=1, 2, ..., M)$,
\begin{equation}\label{eq:RingNeighborsDef}
\rho_i^{\rm ring} \equiv \sum_{\substack{j=1 \\ j\neq i}}^M \Theta \left( 2 \sqrt{\langle R_{g}^2 \rangle} - |{\vec r}_{{\rm CM}, i} - {\vec r}_{{\rm CM}, j}| \right) \, ,
\end{equation}
where $\Theta(x)$ is the Heaviside step function, $\langle R^2_{g} \rangle$ is the mean square gyration radius of the system, and ${\vec r}_{{\rm CM}, j}$  represents the centre of mass position of the $j$-th ring.
According to Eq.~\eqref{eq:RingNeighborsDef}, two rings are defined as ``neighbor'' whenever the spatial distance between their centres of mass is smaller than the twice the root mean-square gyration radius of the system.
We have measured the distribution function of $\rho^{\rm ring}$, $P(\rho^{\rm ring})$ and its mean value, $\langle \rho^{\rm ring} \rangle$, at different confinements and we study these quantities in relation to the distribution of spatial distances between the centres of mass $d_{\rm CM-CM}$ for neighboring rings, $P(d_{\rm CM-CM} \, | \, {\rm neighbors})$. 

\begin{figure}
\includegraphics[width=0.50\textwidth]{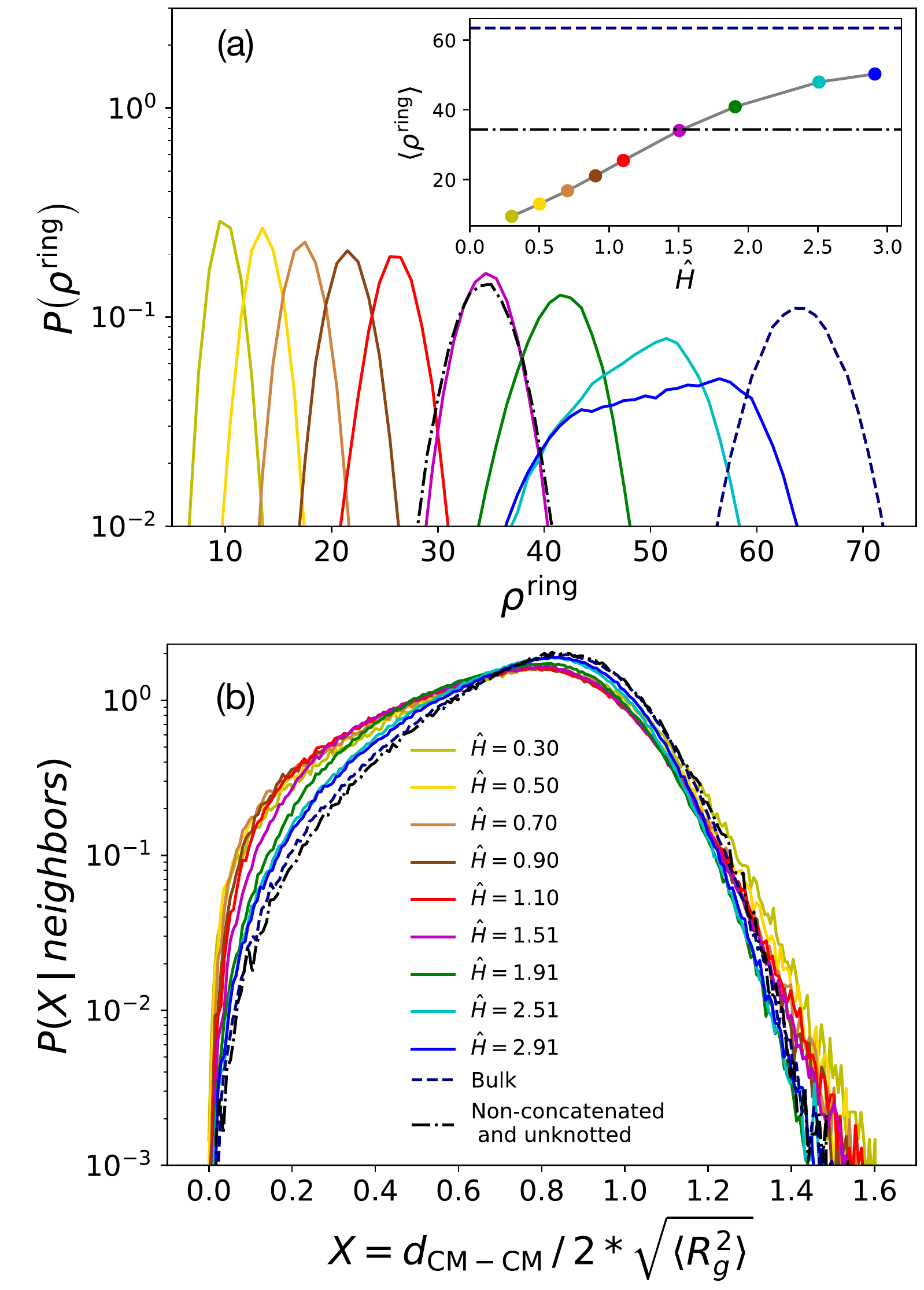}
\caption{
(a)
Distribution function, $P(\rho^{\rm ring})$, of the the number of neighbors per chain $\rho^{\rm ring}$.
Inset: mean number of neighbors per ring, $\langle \rho^{\rm ring} \rangle$.
(b)
Distribution function of the distances between the centres of mass of neighbour chains, $P(d_{\rm CM-CM} \, | \, {\rm neighbours})$, as a function of the variable normalized to twice the root mean-square gyration radius, $2\sqrt{\langle R_g^2 \rangle}$ (Eq.~\eqref{eq:GyrationRadius}), of the rings.
Colors are for different confinements, dashed and dot-dashed lines are for bulk melts and melts of non-concatenated and unknotted rings (see legend).
}
\label{fig:Neighbours}
\end{figure}

Results are shown in Fig.~\ref{fig:Neighbours}, from which it is evident (panel (a)) that $\langle \rho^{\rm ring} \rangle$ decreases as confinement increases, with $\langle \rho^{\rm ring} \rangle$ being always smaller with respect to the bulk reference (dashed line) and even smaller (for the tighter confinements $\hat{\rm H} \lesssim 1.5$) with respect to the non-concatenated and unknotted case (dot-dashed line).
At the same time (panel (b)), the distributions of spatial distances $d_{\rm CM-CM}$ demonstrate that neighboring chains tend to overlap more with each other under stronger confinement. 
Taken all together, we can motivate the reason why the inter-chain contacts decrease in terms of the geometry of the slit.
First, confinement can prevent the formation of stacked conformations along the transverse direction (see Fig.~S3 
in SI), and this surely reduces the inter-chain contacts. 
Moreover, we observe that, by reducing the width of the slit, inter-ring distances tend to increase and this is an effect due to the increasing asymmetry of the slit as confinement increases (see Fig.~S4 
in SI).

\subsubsection{3.2.2. Links}\label{sec:Links}
The reduction of inter-chain contacts should also have consequences on the {\it linking} properties of the confined systems.
To explore this aspect, we adopt the approach developed by us in Ref.~\cite{ubertini2023topological} and compute, $\langle n_{\rm 2link}(|{\rm GLN}|) \rangle$, the mean number of two-chain links at absolute Gauss linking number $|{\rm GLN}|$ and the mean number of distinct three-chain links, $\langle n_{\rm 3link} \rangle$, with given chain topology.

\begin{figure*}
\includegraphics[width=1.0\textwidth]{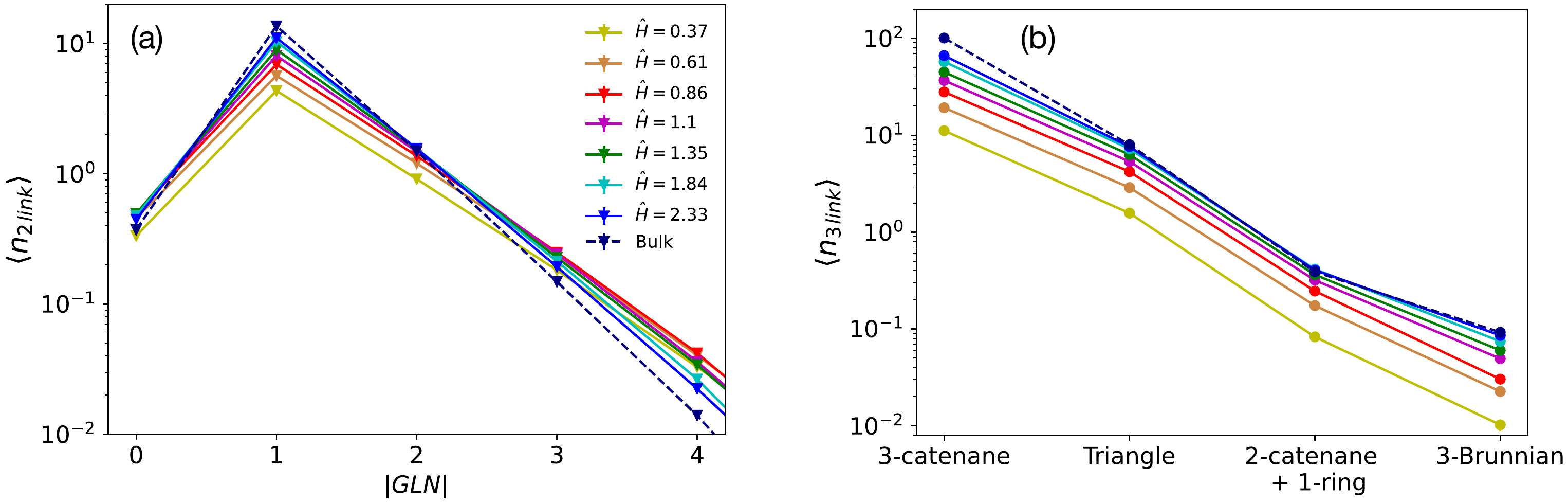}
\caption{
(a)
$\langle n_{\rm 2link} (|{\rm GLN}|) \rangle $, mean number of links per ring with absolute Gauss linking number $|{\rm GLN}|$.
(b)
$\langle n_{\rm 3link}\rangle $, mean number of different three-chain linked structures per ring.
Different colors are for the different confinements, the dashed line is for the bulk system.
}
\label{fig:n-body-links}
\end{figure*}

Results for $\langle n_{\rm 2link}(|{\rm GLN}|) \rangle$ are summarized in panel (a) of Fig.~\ref{fig:n-body-links}.
We notice that ring-ring links are mostly Hopf-like ({\it i.e.}, with $|{\rm GLN}|=1$) and that confinement reduces the extent to which the rings are linked, in agreement with the decrease of overlaps between neighboring chains.
In general, the participation in more complex links decreases exponentially but the rate of decay depends on the level of confinement in the system.
Chains under stronger confinement are characterized by a slower decay, which can be attributed to the fact that neighboring chains penetrate each other more (see Fig.~\ref{fig:Neighbours}(b)).
Additionally, links with $|{\rm GLN}|=0$ ({\it i.e.}, the so-called Whitehead links) have been found between those with $|{\rm GLN}|=2$ and $|{\rm GLN}|=3$ at all confinements.
We further classify these links by computing their Jones polynomial and determining their relative abundances (panel (a) in Fig.~S5 
in SI).
We found that, even in this case, rings under stronger confinement form more complex links with greater ease.

To examine three-chain links, it is necessary to distinguish between two distinct groups of links: those that can be reduced to two-chain links and irreducible ones~\cite{ubertini2023topological}.
The first group include:
(a)
{\it poly(3)catenanes}, chains made of three rings in which two non-concatenated rings are connected to a common ring, and
(b)
{\it triangles}, triplets of rings which are all pairwise concatenated.
Both (a) and (b) can be detected via pairwise linking.
Instead, irreducible three-chain links cannot be detected via pairwise linking and can be further divided in two sub-types:
(c)
{\it poly(2)catenane+1-ring}, structures made of a poly(2)catenane plus another ring which is not directly concatenated (in a pairwise manner) to any of the other two,
and
(d)
{\it Brunnian} links, non-trivial links which become a set of trivial links whenever one component ring is unlinked from the others (the so called {\it Borromean} conformation, the link $6^3_2$, constitutes the easiest example of this kind). 
By resorting to the shrinking method described in~\cite{ubertini2023topological}, we have detected links beloning to the last two classes and computed $\langle n_{\rm 3link}\rangle$ for the different types of three-chain links (Fig.~\ref{fig:n-body-links}(b)).
It is clear from $\langle n_{\rm 3link}\rangle$ that links organize onto a network made almost entirely via pairwise concatenation both in bulk and in confinement.
Irreducible three-chain links are much more rare and decrease with the degree of confinement, for this reason subsequent analysis has been performed by neglecting these three-chain links contribution.
A detailed topological classification of these structures has been reported in Fig.~S5(b) 
in SI, and even in this case three-chain links with higher crossings seem to be more likely for more confined systems. 

\subsubsection{3.2.3. Polymer network and entanglements}\label{sec:PolymerNetwork}
Concatenated rings give rise to a fully connected polymer network~\cite{bobbili2020simulation,ubertini2023topological}.
To characterize this network, we define~\cite{ubertini2023topological} the linking degree ${\rm LD}_i$ of ring $i$,
\begin{equation}\label{eq:MeanLD}
{\rm LD}_i = \sum_{j=1}^M \chi_{ij} \, C_{ij} \, ,
\end{equation}
where the sum runs over the total number of chains in the melt, and where $C_{ij}$ is the $M \times M$ matrix expressing the concatenation status between rings $i$ and $j$:
\begin{equation}\label{eq:Cij-definition}
C_{ij} = \left\{ \begin{array}{cl} 0 \, , & \mbox{if $i=j$} \\ \\ 1 \, , & \mbox{if $i\neq j$ {\it and} form a two-chain link} \\ \\ 0 \, , & \mbox{otherwise} \end{array} \right.
\end{equation}
The ``weight'' factor $\chi_{ij}$ takes into account the ``complexity'' of two-chain links: $\chi_{ij} = |{\rm GLN}|$ or $=\frac{K}2$ depending on whether ${\rm GLN} \neq 0$ or ${\rm GLN} = 0$ respectively.
Here, $K$ is the number of crossings characterizing the link or, in other words, each crossing of the link contributes $1/2$ to an entanglement point.
This quantity is of special interest as we have recently showed~\cite{ubertini2023topological} that the mean value $\langle {\rm LD} \rangle \equiv \langle \frac1M \sum_{i=1}^M {\rm LD}_i \rangle$ is directly connected to the entanglement length of the melt, $N_e$, via the relation $\langle {\rm LD} \rangle = N/N_e$ of the systems. 
To complement this analysis, we have also computed the distribution of the values ${\rm LD}$ at the single ring level, $P({\rm LD})$, which gives us information about the heterogeneity of the network.

\begin{figure}
\includegraphics[width=0.50\textwidth]{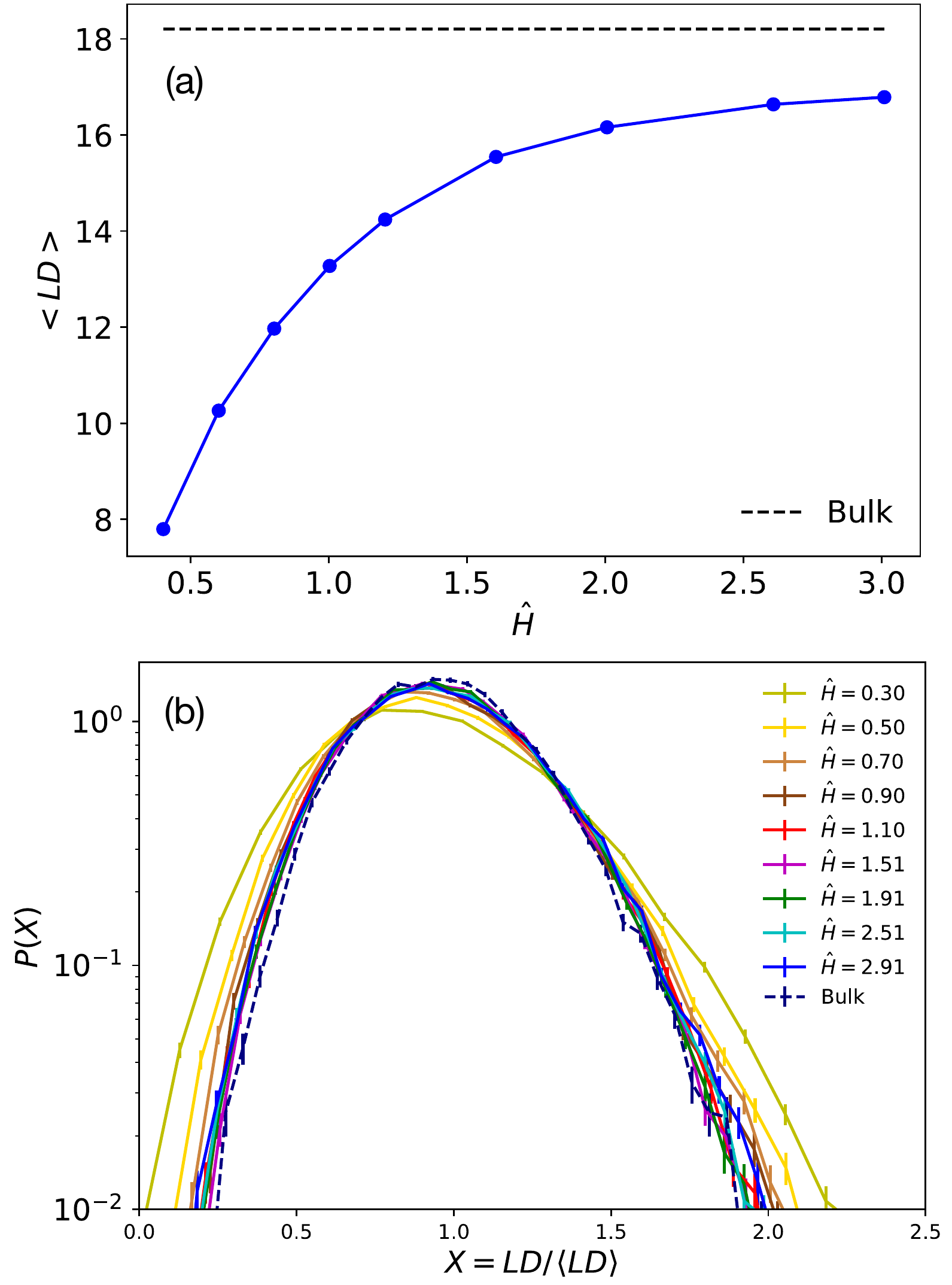}
\caption{
(a)
Mean linking degree, $\langle {\rm LD } \rangle$, as a function of the confinement. The horizontal dotted line represents the bulk value.
(b)
Distribution functions, $P({\rm LD})$, of the linking degree as a function of the variable normalized to the corresponding mean value $\langle {\rm LD} \rangle$.
Different colors are for the different confinements, the dashed line is for the bulk system.
}
\label{fig:LD}
\end{figure}

Results are presented in Fig.~\ref{fig:LD}.
$\langle {\rm LD} \rangle $ (panel (a)) decreases as a function of the confinement, up to a reduction of $\simeq 60\%$ with respect to bulk conditions.
Then, by looking at the distribution functions (panel (b)) of the linking degree as a function of $X={\rm LD} / \langle {\rm LD} \rangle $, we see that the curves at mild confinements display the same behavior for bulk conditions.
Conversely, tails become stronger for more confined systems.
This is in agreement with the behaviour seen for the distribution functions of the sizes of the rings (Fig.~\ref{fig:Rg}(b)), where the tails are higher for stronger confinements.
Fluctuations of ring size may impact on concatenation since smaller rings will be less concatenated having less possibility to reach other rings, while bigger rings can host more contacts and consequently more concatenations.
To sum up, the resulting networks of concatenated rings tend to be more heterogeneous as the confinement become stronger in line with the fluctuations on the rings' sizes.

\section{4. Discussion and  conclusions}\label{sec:DiscConcls}
Our findings illustrate the impact that slit confinement has on the spatial structure of randomly concatenated and knotted ring polymers in melt conditions.

At the single-chain level, our investigation shows that as rings flatten with increasing confinement they tend to adopt more elongated conformations.
At the same time, rings become slightly more rigid with the confinement, a tendency captured by the increase of the correlation ($\langle \cos(\theta) \rangle$, Fig.~\ref{fig:Tangent}(b)) between consecutive bonds along the chain.
We have also demonstrated that the competition between the Kuhn length of the polymers, $\ell_K$, and the height of the slight, ${\rm H}$, induces a non-monotonous behavior on the bond-vector correlation function, $c(\ell)$ (see Fig.~\ref{fig:Tangent}(a, c, d)).
In general, the impact of confinement on ring conformations becomes particularly pronounced with respect to the formation of long intra-chain contacts as the slit narrows (see Fig.~\ref{fig:Pc}), resulting in more compact rings.
Finally, these changes have significant repercussions on the knotting probability which increases with the confinement, and for which we register an increase of $\simeq 130\%$ compared to the bulk value (see Fig.~\ref{fig:Knots}(a)). 

The effects of slit confinement on the inter-chain statistics is similarly noteworthy.
Specifically, as the level of confinement increases, the average number of neighbors per ring, $\langle \rho^{\rm ring} \rangle$, experiences a considerable decrease (see Fig.~\ref{fig:Neighbours}(a)). 
This is directly connected to the decrease of the mean linking degree, $\langle {\rm LD} \rangle$, which displays a total reduction of $\simeq 60\%$ with respect to bulk conditions.\\
\indent This finding has two interesting implications. First, being $\langle {\rm LD} \rangle$ directly related to the mean number of entanglement strands per ring, the decrease of $\langle {\rm LD} \rangle $ means that, at fixed density, confinement alone may alter the entanglement properties of the system making $N_e$ effectively bigger.
This would explains recent findings~\cite{kim2021linear,kim2021ring} showing that for both, linear chains and rings in two-dimensional melts, the resulting dynamical quantities display a quite surprising Rouse-like behavior~\cite{DoiEdwardsBook,RubinsteinColbyBook} which, ultimately, points towards the effective irrelevance of entanglement effects due to inter-chain interactions.
It is worth stressing that, although the mean number of ring-ring concatenations ({\it i.e.}, $\langle {\rm LD} \rangle$) decreasing with confinement is not entirely surprisingly (in $2d$ rings can not be concatenated), the important point to stress here is that the works~\cite{kim2021linear,kim2021ring} and our analysis here and in~\cite{ubertini2023topological} suggest that entanglements are indeed well captured~\cite{qin2014tubes,milner2020unified,bobbili2020simulation} by two-chain topological links alone. Finally, it is worth recalling that the elastic plateau modulus $G_0$, which quantifies the stress-strain relationship of polymeric materials, is related to the total number of entanglement strands of the melt, $G_0 \propto \frac{N M}{N_e}$~\cite{RubinsteinColbyBook}.
Then our results imply that, as confinement grows, the resulting network becomes softer ($G_0$ decreases), highlighting the important role of geometric confinement on the mechanical properties of the stored polymer network.

\section{Supporting Information}
Time mean-square displacement of monomers in the frame of the centre of mass of the corresponding ring,
fractional population of knot types,
contour plots for the joint distribution function of parallel and transverse components of the distances between the centres of mass of neighboring rings,
distribution functions of the distances between the rings' centres of mass,
fractional population of two-chain links with ${\rm GLN}=0$.

\section{Acknowledgement}
The authors acknowledge networking support by the COST Action CA17139 (EUTOPIA). 
\bibliography{biblio}

\begin{thebibliography}{33}
\expandafter\ifx\csname natexlab\endcsname\relax\def\natexlab#1{#1}\fi
\expandafter\ifx\csname bibnamefont\endcsname\relax
  \def\bibnamefont#1{#1}\fi
\expandafter\ifx\csname bibfnamefont\endcsname\relax
  \def\bibfnamefont#1{#1}\fi
\expandafter\ifx\csname citenamefont\endcsname\relax
  \def\citenamefont#1{#1}\fi
\expandafter\ifx\csname url\endcsname\relax
  \def\url#1{\texttt{#1}}\fi
\expandafter\ifx\csname urlprefix\endcsname\relax\def\urlprefix{URL }\fi
\providecommand{\bibinfo}[2]{#2}
\providecommand{\eprint}[2][]{\url{#2}}

\bibitem[{\citenamefont{Wu et~al.}(2017)\citenamefont{Wu, Rauscher, Lang,
  Wojtecki, De~Pablo, Hore, and Rowan}}]{wu2017poly}
\bibinfo{author}{\bibfnamefont{Q.}~\bibnamefont{Wu}},
  \bibinfo{author}{\bibfnamefont{P.~M.} \bibnamefont{Rauscher}},
  \bibinfo{author}{\bibfnamefont{X.}~\bibnamefont{Lang}},
  \bibinfo{author}{\bibfnamefont{R.~J.} \bibnamefont{Wojtecki}},
  \bibinfo{author}{\bibfnamefont{J.~J.} \bibnamefont{De~Pablo}},
  \bibinfo{author}{\bibfnamefont{M.~J.} \bibnamefont{Hore}}, \bibnamefont{and}
  \bibinfo{author}{\bibfnamefont{S.~J.} \bibnamefont{Rowan}},
  \bibinfo{journal}{Science} \textbf{\bibinfo{volume}{358}},
  \bibinfo{pages}{1434} (\bibinfo{year}{2017}).

\bibitem[{\citenamefont{Hart et~al.}(2021)\citenamefont{Hart, Hertzog,
  Rauscher, Rawe, Tranquilli, and Rowan}}]{hart2021material}
\bibinfo{author}{\bibfnamefont{L.~F.} \bibnamefont{Hart}},
  \bibinfo{author}{\bibfnamefont{J.~E.} \bibnamefont{Hertzog}},
  \bibinfo{author}{\bibfnamefont{P.~M.} \bibnamefont{Rauscher}},
  \bibinfo{author}{\bibfnamefont{B.~W.} \bibnamefont{Rawe}},
  \bibinfo{author}{\bibfnamefont{M.~M.} \bibnamefont{Tranquilli}},
  \bibnamefont{and} \bibinfo{author}{\bibfnamefont{S.~J.} \bibnamefont{Rowan}},
  \bibinfo{journal}{Nature Reviews Materials} \textbf{\bibinfo{volume}{6}},
  \bibinfo{pages}{508} (\bibinfo{year}{2021}).

\bibitem[{\citenamefont{Krajina et~al.}(2018)\citenamefont{Krajina, Zhu,
  Heilshorn, and Spakowitz}}]{krajina2018active}
\bibinfo{author}{\bibfnamefont{B.~A.} \bibnamefont{Krajina}},
  \bibinfo{author}{\bibfnamefont{A.}~\bibnamefont{Zhu}},
  \bibinfo{author}{\bibfnamefont{S.~C.} \bibnamefont{Heilshorn}},
  \bibnamefont{and} \bibinfo{author}{\bibfnamefont{A.~J.}
  \bibnamefont{Spakowitz}}, \bibinfo{journal}{Phys. Rev. Lett.}
  \textbf{\bibinfo{volume}{121}}, \bibinfo{pages}{148001}
  (\bibinfo{year}{2018}).

\bibitem[{\citenamefont{Rapha{\"e}l et~al.}(1997)\citenamefont{Rapha{\"e}l,
  Gay, and de~Gennes}}]{deGennes1997}
\bibinfo{author}{\bibfnamefont{E.}~\bibnamefont{Rapha{\"e}l}},
  \bibinfo{author}{\bibfnamefont{C.}~\bibnamefont{Gay}}, \bibnamefont{and}
  \bibinfo{author}{\bibfnamefont{P.~G.} \bibnamefont{de~Gennes}},
  \bibinfo{journal}{Journal of Statistical Physics}
  \textbf{\bibinfo{volume}{89}}, \bibinfo{pages}{111} (\bibinfo{year}{1997}).

\bibitem[{\citenamefont{Chen et~al.}(1995)\citenamefont{Chen, Rauch, White,
  Englund, and Cozzarelli}}]{chen1995topology}
\bibinfo{author}{\bibfnamefont{J.}~\bibnamefont{Chen}},
  \bibinfo{author}{\bibfnamefont{C.~A.} \bibnamefont{Rauch}},
  \bibinfo{author}{\bibfnamefont{J.~H.} \bibnamefont{White}},
  \bibinfo{author}{\bibfnamefont{P.~T.} \bibnamefont{Englund}},
  \bibnamefont{and} \bibinfo{author}{\bibfnamefont{N.~R.}
  \bibnamefont{Cozzarelli}}, \bibinfo{journal}{Cell}
  \textbf{\bibinfo{volume}{80}}, \bibinfo{pages}{61} (\bibinfo{year}{1995}).

\bibitem[{\citenamefont{Chiarantoni and
  Micheletti}(2022)}]{chiarantoni2022effect}
\bibinfo{author}{\bibfnamefont{P.}~\bibnamefont{Chiarantoni}} \bibnamefont{and}
  \bibinfo{author}{\bibfnamefont{C.}~\bibnamefont{Micheletti}},
  \bibinfo{journal}{Macromolecules} \textbf{\bibinfo{volume}{55}},
  \bibinfo{pages}{4523} (\bibinfo{year}{2022}).

\bibitem[{\citenamefont{Doi and Edwards}(1986)}]{DoiEdwardsBook}
\bibinfo{author}{\bibfnamefont{M.}~\bibnamefont{Doi}} \bibnamefont{and}
  \bibinfo{author}{\bibfnamefont{S.~F.} \bibnamefont{Edwards}},
  \emph{\bibinfo{title}{The Theory of Polymer Dynamics}}
  (\bibinfo{publisher}{Clarendon}, \bibinfo{address}{Oxford},
  \bibinfo{year}{1986}).

\bibitem[{\citenamefont{Rubinstein and Colby}(2003)}]{RubinsteinColbyBook}
\bibinfo{author}{\bibfnamefont{M.}~\bibnamefont{Rubinstein}} \bibnamefont{and}
  \bibinfo{author}{\bibfnamefont{R.~H.} \bibnamefont{Colby}},
  \emph{\bibinfo{title}{Polymer Physics}} (\bibinfo{publisher}{Oxford
  University Press}, \bibinfo{address}{New York}, \bibinfo{year}{2003}).

\bibitem[{\citenamefont{Rauscher
  et~al.}(2020{\natexlab{a}})\citenamefont{Rauscher, Schweizer, Rowan, and
  de~Pablo}}]{rauscher2020dynamics}
\bibinfo{author}{\bibfnamefont{P.~M.} \bibnamefont{Rauscher}},
  \bibinfo{author}{\bibfnamefont{K.~S.} \bibnamefont{Schweizer}},
  \bibinfo{author}{\bibfnamefont{S.~J.} \bibnamefont{Rowan}}, \bibnamefont{and}
  \bibinfo{author}{\bibfnamefont{J.~J.} \bibnamefont{de~Pablo}},
  \bibinfo{journal}{The Journal of Chemical Physics}
  \textbf{\bibinfo{volume}{152}}, \bibinfo{pages}{214901}
  (\bibinfo{year}{2020}{\natexlab{a}}).

\bibitem[{\citenamefont{Rauscher
  et~al.}(2020{\natexlab{b}})\citenamefont{Rauscher, Schweizer, Rowan, and
  De~Pablo}}]{rauscher2020thermodynamics}
\bibinfo{author}{\bibfnamefont{P.~M.} \bibnamefont{Rauscher}},
  \bibinfo{author}{\bibfnamefont{K.~S.} \bibnamefont{Schweizer}},
  \bibinfo{author}{\bibfnamefont{S.~J.} \bibnamefont{Rowan}}, \bibnamefont{and}
  \bibinfo{author}{\bibfnamefont{J.~J.} \bibnamefont{De~Pablo}},
  \bibinfo{journal}{Macromolecules} \textbf{\bibinfo{volume}{53}},
  \bibinfo{pages}{3390} (\bibinfo{year}{2020}{\natexlab{b}}).

\bibitem[{\citenamefont{Lang et~al.}(2012)\citenamefont{Lang, Fischer, and
  Sommer}}]{lang2012effect}
\bibinfo{author}{\bibfnamefont{M.}~\bibnamefont{Lang}},
  \bibinfo{author}{\bibfnamefont{J.}~\bibnamefont{Fischer}}, \bibnamefont{and}
  \bibinfo{author}{\bibfnamefont{J.-U.} \bibnamefont{Sommer}},
  \bibinfo{journal}{Macromolecules} \textbf{\bibinfo{volume}{45}},
  \bibinfo{pages}{7642} (\bibinfo{year}{2012}).

\bibitem[{\citenamefont{Lang et~al.}(2014)\citenamefont{Lang, Fischer, Werner,
  and Sommer}}]{lang2014swelling}
\bibinfo{author}{\bibfnamefont{M.}~\bibnamefont{Lang}},
  \bibinfo{author}{\bibfnamefont{J.}~\bibnamefont{Fischer}},
  \bibinfo{author}{\bibfnamefont{M.}~\bibnamefont{Werner}}, \bibnamefont{and}
  \bibinfo{author}{\bibfnamefont{J.-U.} \bibnamefont{Sommer}},
  \bibinfo{journal}{Phys. Rev. Lett.} \textbf{\bibinfo{volume}{112}},
  \bibinfo{pages}{238001} (\bibinfo{year}{2014}).

\bibitem[{\citenamefont{Ubertini and Rosa}(2021)}]{ubertini2021computer}
\bibinfo{author}{\bibfnamefont{M.~A.} \bibnamefont{Ubertini}} \bibnamefont{and}
  \bibinfo{author}{\bibfnamefont{A.}~\bibnamefont{Rosa}},
  \bibinfo{journal}{Phys. Rev. E} \textbf{\bibinfo{volume}{104}},
  \bibinfo{pages}{054503} (\bibinfo{year}{2021}).

\bibitem[{\citenamefont{Ubertini and Rosa}(2023)}]{ubertini2023topological}
\bibinfo{author}{\bibfnamefont{M.~A.} \bibnamefont{Ubertini}} \bibnamefont{and}
  \bibinfo{author}{\bibfnamefont{A.}~\bibnamefont{Rosa}},
  \bibinfo{journal}{Macromolecules} \textbf{\bibinfo{volume}{56}},
  \bibinfo{pages}{3354} (\bibinfo{year}{2023}).

\bibitem[{\citenamefont{Soh and Doyle}(2021)}]{soh2021equilibrium}
\bibinfo{author}{\bibfnamefont{B.~W.} \bibnamefont{Soh}} \bibnamefont{and}
  \bibinfo{author}{\bibfnamefont{P.~S.} \bibnamefont{Doyle}},
  \bibinfo{journal}{ACS Macro Letters} \textbf{\bibinfo{volume}{10}},
  \bibinfo{pages}{880} (\bibinfo{year}{2021}).

\bibitem[{\citenamefont{Ubertini et~al.}(2022)\citenamefont{Ubertini, Smrek,
  and Rosa}}]{ubertini2022double}
\bibinfo{author}{\bibfnamefont{M.~A.} \bibnamefont{Ubertini}},
  \bibinfo{author}{\bibfnamefont{J.}~\bibnamefont{Smrek}}, \bibnamefont{and}
  \bibinfo{author}{\bibfnamefont{A.}~\bibnamefont{Rosa}},
  \bibinfo{journal}{Macromolecules} \textbf{\bibinfo{volume}{55}},
  \bibinfo{pages}{10723} (\bibinfo{year}{2022}).

\bibitem[{Mea()}]{MeanBondLengthNote}
\bibinfo{note}{Notice that the values for $\langle b \rangle$ between the bulk
  and the confined rings are all very close to each other, with a slight
  increase as confinement decreases.}

\bibitem[{\citenamefont{Kremer and Grest}(1990)}]{KremerGrest1992}
\bibinfo{author}{\bibfnamefont{K.}~\bibnamefont{Kremer}} \bibnamefont{and}
  \bibinfo{author}{\bibfnamefont{G.~S.} \bibnamefont{Grest}},
  \bibinfo{journal}{The Journal of Chemical Physics}
  \textbf{\bibinfo{volume}{92}}, \bibinfo{pages}{5057} (\bibinfo{year}{1990}).

\bibitem[{\citenamefont{Jones}(1985)}]{Jones1985}
\bibinfo{author}{\bibfnamefont{V.~F.~R.} \bibnamefont{Jones}},
  \bibinfo{journal}{Bulletin of the American Mathematical Society}
  \textbf{\bibinfo{volume}{12}}, \bibinfo{pages}{103} (\bibinfo{year}{1985}).

\bibitem[{\citenamefont{Dabrowski-Tumanski
  et~al.}(2021)\citenamefont{Dabrowski-Tumanski, Rubach, Niemyska, Gren, and
  Sulkowska}}]{dabrowski2021topoly}
\bibinfo{author}{\bibfnamefont{P.}~\bibnamefont{Dabrowski-Tumanski}},
  \bibinfo{author}{\bibfnamefont{P.}~\bibnamefont{Rubach}},
  \bibinfo{author}{\bibfnamefont{W.}~\bibnamefont{Niemyska}},
  \bibinfo{author}{\bibfnamefont{B.~A.} \bibnamefont{Gren}}, \bibnamefont{and}
  \bibinfo{author}{\bibfnamefont{J.~I.} \bibnamefont{Sulkowska}},
  \bibinfo{journal}{Briefings in Bioinformatics} \textbf{\bibinfo{volume}{22}},
  \bibinfo{pages}{bbaa196} (\bibinfo{year}{2021}).

\bibitem[{\citenamefont{Rolfsen}(2003)}]{Rolfsen2003KnotsAL}
\bibinfo{author}{\bibfnamefont{D.}~\bibnamefont{Rolfsen}},
  \emph{\bibinfo{title}{Knots and links}} (\bibinfo{publisher}{AMS Chelsea
  Publishing}, \bibinfo{year}{2003}).

\bibitem[{\citenamefont{D’Adamo et~al.}(2017)\citenamefont{D’Adamo,
  Orlandini, and Micheletti}}]{d2017linking}
\bibinfo{author}{\bibfnamefont{G.}~\bibnamefont{D’Adamo}},
  \bibinfo{author}{\bibfnamefont{E.}~\bibnamefont{Orlandini}},
  \bibnamefont{and}
  \bibinfo{author}{\bibfnamefont{C.}~\bibnamefont{Micheletti}},
  \bibinfo{journal}{Macromolecules} \textbf{\bibinfo{volume}{50}},
  \bibinfo{pages}{1713} (\bibinfo{year}{2017}).

\bibitem[{\citenamefont{Rosa and Everaers}(2019)}]{rosa2019conformational}
\bibinfo{author}{\bibfnamefont{A.}~\bibnamefont{Rosa}} \bibnamefont{and}
  \bibinfo{author}{\bibfnamefont{R.}~\bibnamefont{Everaers}},
  \bibinfo{journal}{The European Physical Journal E}
  \textbf{\bibinfo{volume}{42}}, \bibinfo{pages}{1} (\bibinfo{year}{2019}).

\bibitem[{\citenamefont{Halverson et~al.}(2014)\citenamefont{Halverson, Smrek,
  Kremer, and Grosberg}}]{halverson2014melt}
\bibinfo{author}{\bibfnamefont{J.~D.} \bibnamefont{Halverson}},
  \bibinfo{author}{\bibfnamefont{J.}~\bibnamefont{Smrek}},
  \bibinfo{author}{\bibfnamefont{K.}~\bibnamefont{Kremer}}, \bibnamefont{and}
  \bibinfo{author}{\bibfnamefont{A.~Y.} \bibnamefont{Grosberg}},
  \bibinfo{journal}{Reports on Progress in Physics}
  \textbf{\bibinfo{volume}{77}}, \bibinfo{pages}{022601}
  (\bibinfo{year}{2014}).

\bibitem[{\citenamefont{Tesi et~al.}(1994)\citenamefont{Tesi, van Rensburgs,
  Orlandini, and Whittington}}]{tesi1994knot}
\bibinfo{author}{\bibfnamefont{M.}~\bibnamefont{Tesi}},
  \bibinfo{author}{\bibfnamefont{E.~J.} \bibnamefont{van Rensburgs}},
  \bibinfo{author}{\bibfnamefont{E.}~\bibnamefont{Orlandini}},
  \bibnamefont{and}
  \bibinfo{author}{\bibfnamefont{S.}~\bibnamefont{Whittington}},
  \bibinfo{journal}{Journal of Physics A: Mathematical and General}
  \textbf{\bibinfo{volume}{27}}, \bibinfo{pages}{347} (\bibinfo{year}{1994}).

\bibitem[{\citenamefont{Micheletti et~al.}(2011)\citenamefont{Micheletti,
  Marenduzzo, and Orlandini}}]{micheletti2011polymers}
\bibinfo{author}{\bibfnamefont{C.}~\bibnamefont{Micheletti}},
  \bibinfo{author}{\bibfnamefont{D.}~\bibnamefont{Marenduzzo}},
  \bibnamefont{and}
  \bibinfo{author}{\bibfnamefont{E.}~\bibnamefont{Orlandini}},
  \bibinfo{journal}{Physics Reports} \textbf{\bibinfo{volume}{504}},
  \bibinfo{pages}{1} (\bibinfo{year}{2011}).

\bibitem[{\citenamefont{Micheletti and
  Orlandini}(2012)}]{micheletti2012numerical}
\bibinfo{author}{\bibfnamefont{C.}~\bibnamefont{Micheletti}} \bibnamefont{and}
  \bibinfo{author}{\bibfnamefont{E.}~\bibnamefont{Orlandini}},
  \bibinfo{journal}{Macromolecules} \textbf{\bibinfo{volume}{45}},
  \bibinfo{pages}{2113} (\bibinfo{year}{2012}).

\bibitem[{\citenamefont{Bobbili and Milner}(2020)}]{bobbili2020simulation}
\bibinfo{author}{\bibfnamefont{S.~V.} \bibnamefont{Bobbili}} \bibnamefont{and}
  \bibinfo{author}{\bibfnamefont{S.~T.} \bibnamefont{Milner}},
  \bibinfo{journal}{Macromolecules} \textbf{\bibinfo{volume}{53}},
  \bibinfo{pages}{3861} (\bibinfo{year}{2020}).

\bibitem[{\citenamefont{Kim et~al.}(2021{\natexlab{a}})\citenamefont{Kim, Kim,
  and Baig}}]{kim2021linear}
\bibinfo{author}{\bibfnamefont{J.}~\bibnamefont{Kim}},
  \bibinfo{author}{\bibfnamefont{J.~M.} \bibnamefont{Kim}}, \bibnamefont{and}
  \bibinfo{author}{\bibfnamefont{C.}~\bibnamefont{Baig}},
  \bibinfo{journal}{Polymer} \textbf{\bibinfo{volume}{213}},
  \bibinfo{pages}{123308} (\bibinfo{year}{2021}{\natexlab{a}}).

\bibitem[{\citenamefont{Kim et~al.}(2021{\natexlab{b}})\citenamefont{Kim, Kim,
  and Baig}}]{kim2021ring}
\bibinfo{author}{\bibfnamefont{J.}~\bibnamefont{Kim}},
  \bibinfo{author}{\bibfnamefont{J.~M.} \bibnamefont{Kim}}, \bibnamefont{and}
  \bibinfo{author}{\bibfnamefont{C.}~\bibnamefont{Baig}},
  \bibinfo{journal}{Soft Matter} \textbf{\bibinfo{volume}{17}},
  \bibinfo{pages}{10703} (\bibinfo{year}{2021}{\natexlab{b}}).

\bibitem[{\citenamefont{Qin and Milner}(2014)}]{qin2014tubes}
\bibinfo{author}{\bibfnamefont{J.}~\bibnamefont{Qin}} \bibnamefont{and}
  \bibinfo{author}{\bibfnamefont{S.~T.} \bibnamefont{Milner}},
  \bibinfo{journal}{Macromolecules} \textbf{\bibinfo{volume}{47}},
  \bibinfo{pages}{6077} (\bibinfo{year}{2014}).

\bibitem[{\citenamefont{Milner}(2020)}]{milner2020unified}
\bibinfo{author}{\bibfnamefont{S.~T.} \bibnamefont{Milner}},
  \bibinfo{journal}{Macromolecules} \textbf{\bibinfo{volume}{53}},
  \bibinfo{pages}{1314} (\bibinfo{year}{2020}).

\bibitem[{\citenamefont{Hoste et~al.}(1998)\citenamefont{Hoste, Thistlethwaite,
  and Weeks}}]{hoste1998first}
\bibinfo{author}{\bibfnamefont{J.}~\bibnamefont{Hoste}},
  \bibinfo{author}{\bibfnamefont{M.}~\bibnamefont{Thistlethwaite}},
  \bibnamefont{and} \bibinfo{author}{\bibfnamefont{J.}~\bibnamefont{Weeks}},
  \bibinfo{journal}{Math. Intelligencer} \textbf{\bibinfo{volume}{20}},
  \bibinfo{pages}{33} (\bibinfo{year}{1998}).

\end{thebibliography}

\clearpage

\widetext
\clearpage
\begin{center}
\textbf{
-- Supporting Information -- \\ \vspace*{1.5mm}
Spatial organization of slit-confined melts of ring polymers with non-conserved topology: A lattice Monte Carlo study
}
\\
\vspace*{5mm}
Mattia Alberto Ubertini and Angelo Rosa
\vspace*{10mm}
\end{center}
\balancecolsandclearpage

\setcounter{equation}{0}
\setcounter{figure}{0}
\setcounter{table}{0}
\setcounter{page}{1}
\setcounter{section}{0}
\makeatletter
\renewcommand{\theequation}{S\arabic{equation}}
\renewcommand{\thefigure}{S\arabic{figure}}
\renewcommand{\thetable}{S\arabic{table}}
\renewcommand{\thesection}{S\arabic{section}}

\makeatletter
\@fpsep\textheight
\makeatother

\begin{figure*}
\includegraphics[width=0.5\textwidth]{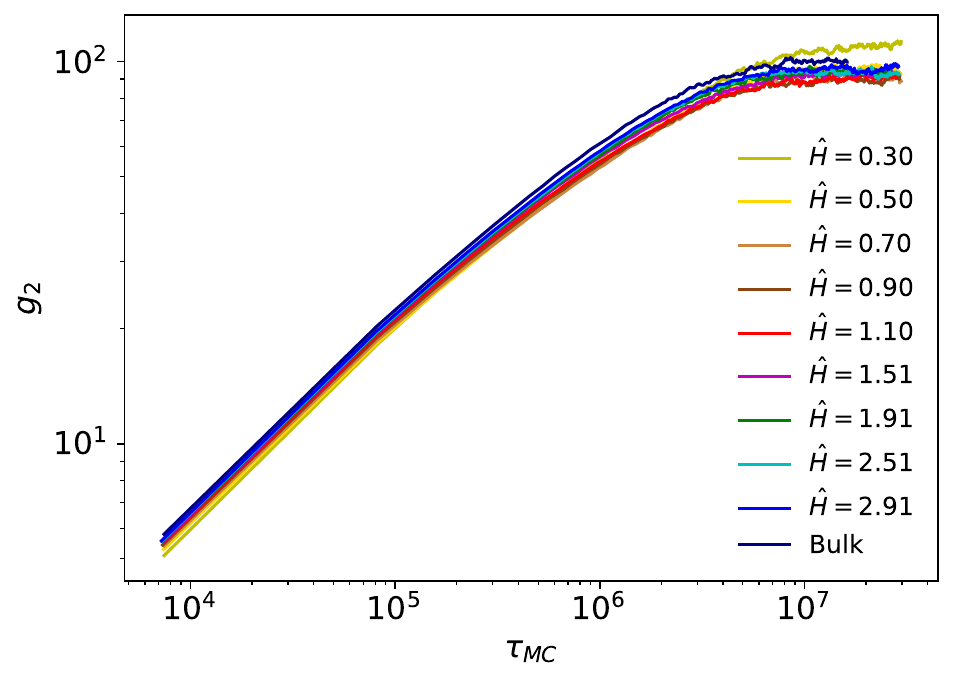}
\caption{
Time mean-square displacement of monomers in the centre of mass of the corresponding ring~\cite{KremerGrest1992,ubertini2021computer},
\\
\\
$
\mbox{\hspace{4.0cm}}
g_2(\tau_{\rm MC}) \equiv \left\langle \frac1M \sum_{m=1}^M \left[ ({\vec r}_m(\tau_{\rm MC}) - {\vec r}_{\rm CM}(\tau_{\rm MC})) - ({\vec r}_m(0) - {\vec r}_{\rm CM}(0)) \right]^2 \right\rangle \, ,
$
\\
\\
as a function of the Monte Carlo (MC) time, $\tau_{\rm MC}$.
$\vec r_m$ is the spatial position of monomer $m$, while $\vec r_{\rm CM}$ is the spatial position of the centre of mass of the ring. 
All examined systems display a plateau, that is indicative of their successful equilibration. 
The time scale associated to the onset of the plateau is used to estimate which amount from the initial portion of the corresponding MC trajectory has to be discarded in order to compute rings' properties accurately.
}
\label{fig:g2}
\end{figure*}
\begin{figure*}
\includegraphics[width=1\textwidth]{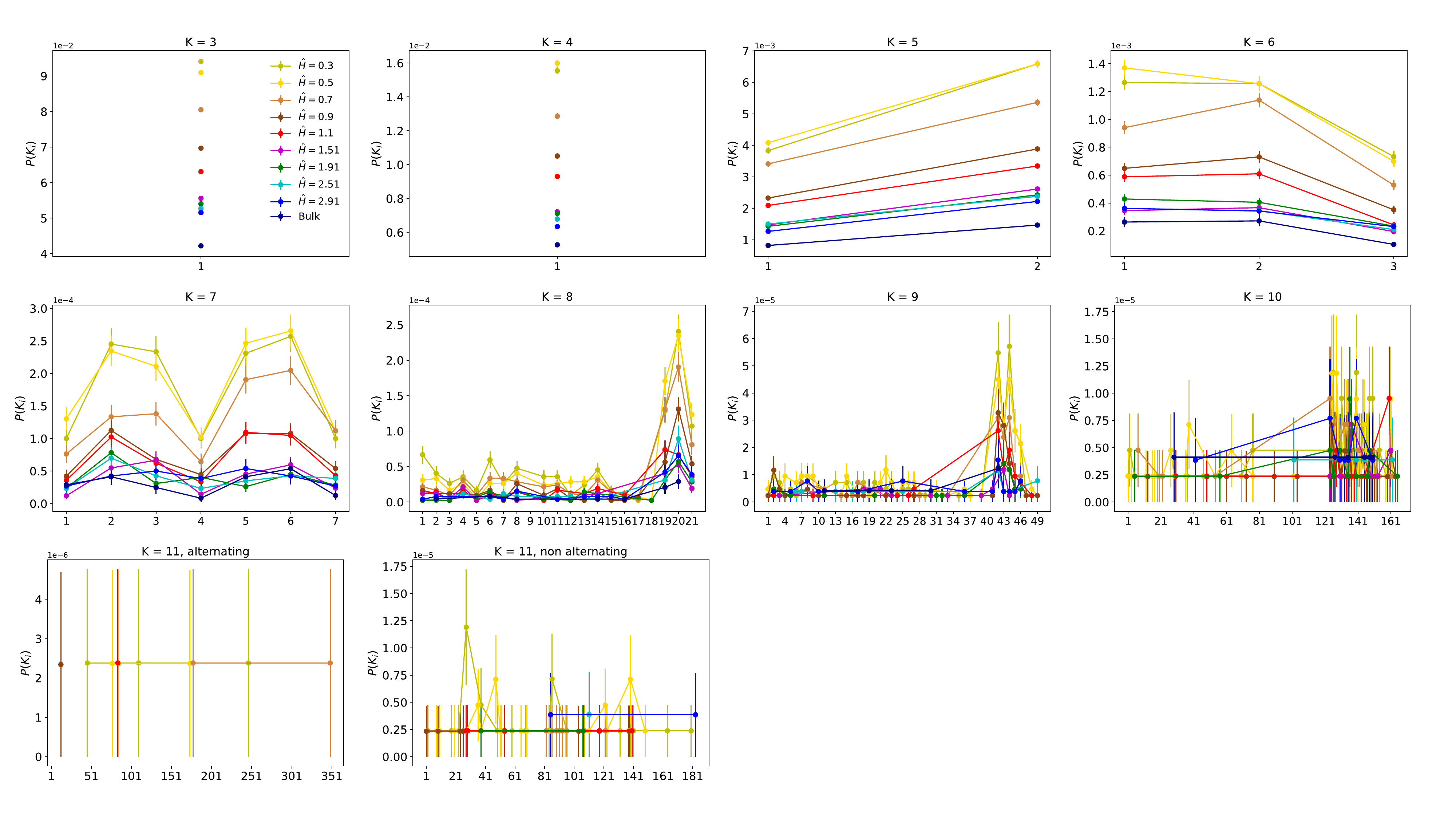}
\caption{
Fractional population of each knot type -- named according to the Rolfsen convention~\cite{Rolfsen2003KnotsAL} -- at fixed number of crossings $K$ and for the different values of confinement $\hat{\text{H}}$ and in bulk conditions (see legend).
For $K=11$, knots are categorized according the Hoste-Thistlethwaite table~\cite{hoste1998first} that split them in {\it alternating} ($K_{a\_i}$) and {\it non-alternating} ones ($K_{n\_i}$), with the index $i$ used to enumerate the rings separately within each group. 
Large relative error bars are due to the limited size of the sample (notice that the values of each $y$-axis have to be multiplied by the power-law reported on the top left corner of the corresponding panel).
}
\label{fig:Distribution_Knots}
\end{figure*}
\begin{figure*}
\includegraphics[width=1\textwidth]{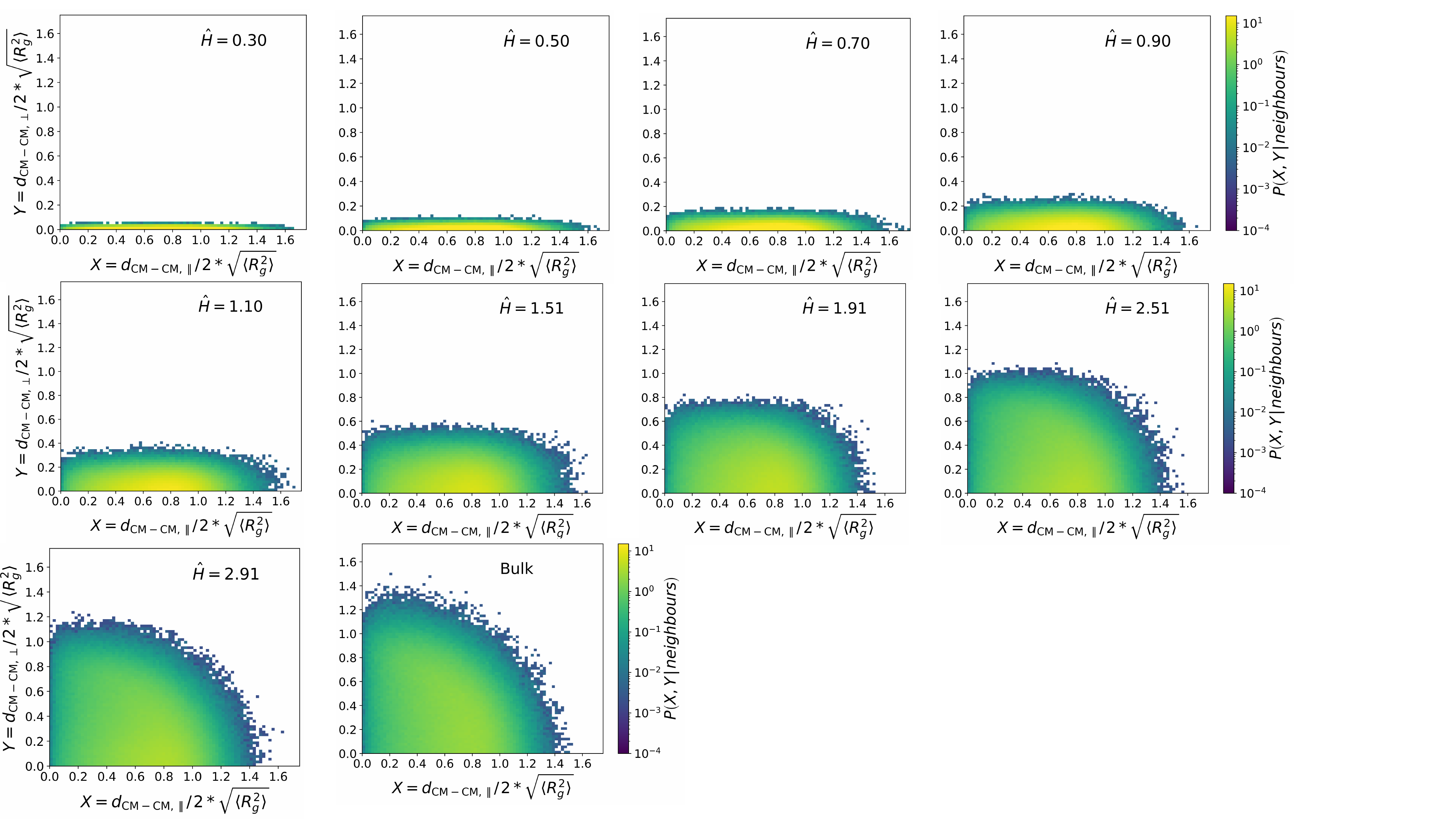}
\caption{
Contour plots for the joint distribution function of the slit-parallel and slit-transverse (or, slit-perpendicular) components of the distances between the centres of mass of neighboring rings
$P(d_{{\rm CM}-{\rm CM}, \parallel}, d_{{\rm CM}-{\rm CM}, \perp} \, | \, {\rm neighbors})$.
Distances have been rescaled by the corresponding root-mean-square gyration radius of the rings.
}
\label{fig:P-2d}
\end{figure*}
\begin{figure*}
\includegraphics[width=0.75\textwidth]{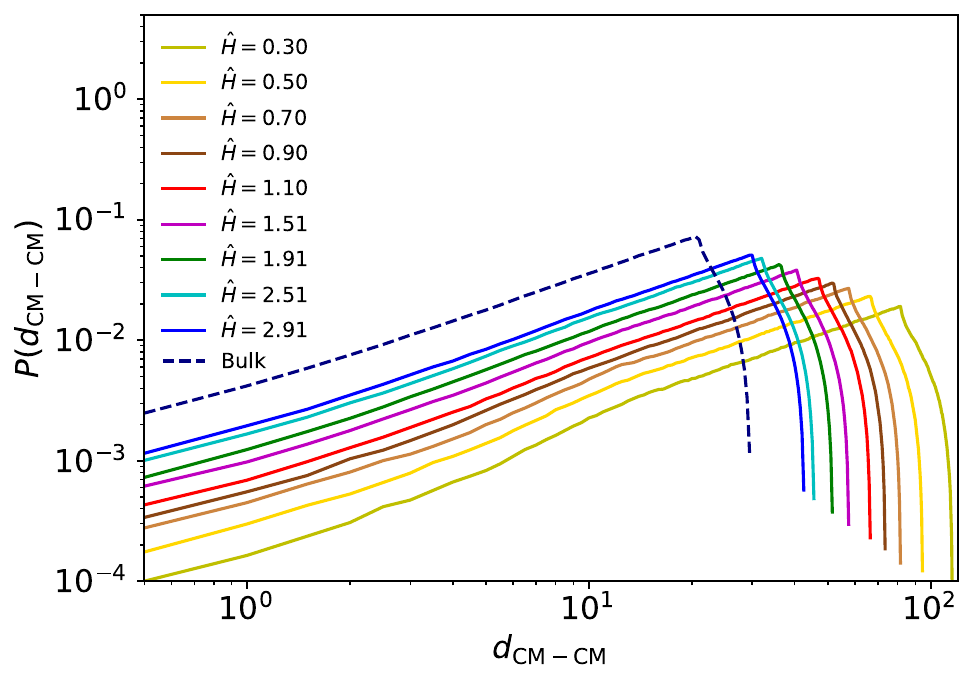}
\caption{
Distribution functions of the distances between the rings' centres of mass, $P(d_{{\rm CM}-{\rm CM}})$.}
\label{fig:distance_all_rigs}
\end{figure*}
\begin{figure*}
\includegraphics[width=1\textwidth]{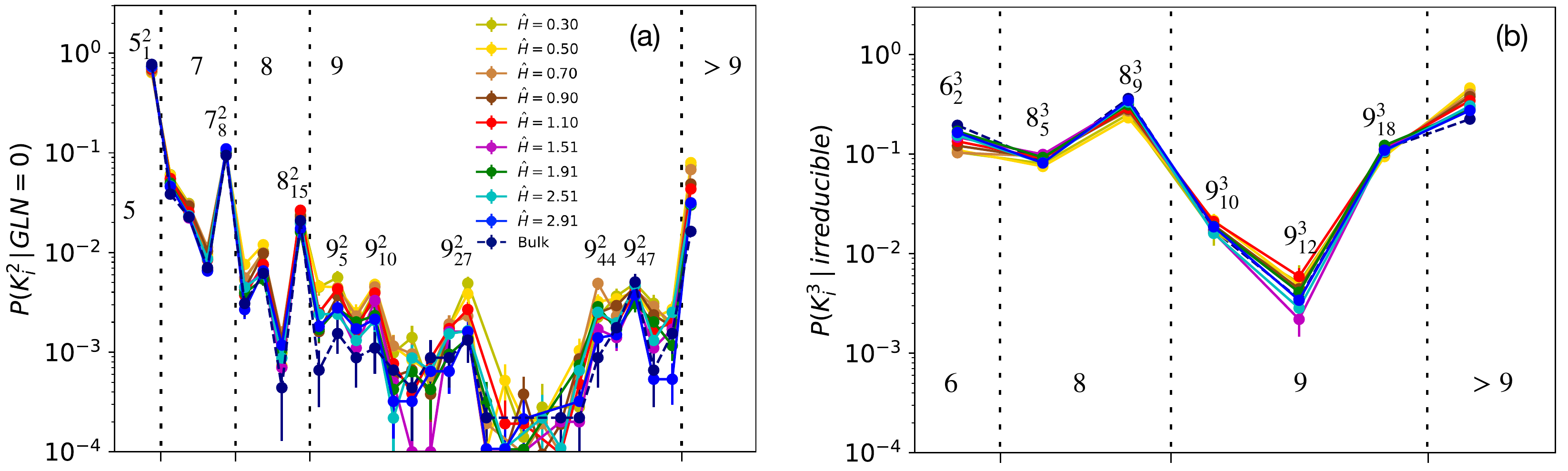}
\caption{
(a)
$P(K_i^2 | {\rm GLN}=0)$, fractional population of two-chain links $K_i^2$ (termed according to the Rolfsen convention~\cite{Rolfsen2003KnotsAL}) having ${\rm GLN} = 0$.
Labels have been put for links corresponding to peaks of the distributions. 
(b)
$P(K_i^3 | \mbox{irreducible})$, fractional population of three-chain links $K_i^3$ (also termed according to the Rolfsen convention~\cite{Rolfsen2003KnotsAL}) belonging to the poly(2)catenane+1-ring and Brunnian classes (see main text for details).
{\it Topoly} fails~\cite{dabrowski2021topoly} recognizing links with $>9$ crossings, so these lack categorization. 
In both panels, vertical dotted lines delimit areas of links at fixed number of crossings $K$.
}
\label{fig:Statistics_Links}
\end{figure*}
\end{document}